\documentclass{svjour3}
\usepackage{natbib}
\usepackage{color}
\usepackage{times}
\usepackage{latexsym}
\usepackage{enumerate}
\usepackage{epsfig}
\usepackage{amsmath}
\usepackage{amssymb}
\usepackage{amsfonts}
\usepackage[english]{babel}
\usepackage{graphicx}
\usepackage{graphics}
\usepackage{psfrag}
\usepackage{amsbsy}
\usepackage{amsmath}

\def\bb{\mathbf{b}}
\def\bx{\mathbf{x}}
\def\by{\mathbf{y}}
\def\bz{\mathbf{z}}

\def\bt{\mathbf{t}}
\def\bw{\mbox{\bf w}}

\def\bT{\mathbf{T}}
\def\bX{\mathbf{X}}

\def\bW{\mathbf{W}}

\def\bZ{\mathbf{Z}}
\def\b0{\mathbf{0}}
\def\b1{\mathbf{1}}

\def\cD{\mathcal{D}}
\def\cE{\mathcal{E}}

\def\cO{\mathcal{O}}
\def\cX{\mathcal{X}}
\def\cY{\mathcal{Y}}

\def\cZ{\mathcal{Z}}
\def\mR{\mathbb{R}}

\def\EE{\mathbb{E}}

\def\bmu{\mbox{\boldmath $\mu$}}

\def\bpsi{\mbox{\boldmath $\psi$}}

\def\bbeta{\mbox{\boldmath $\beta$}}

\def\bxi{\mbox{\boldmath $\xi$}}
\def\btheta{\mbox{\boldmath $\theta$}}

\def\bSigma{\mbox{\boldmath $\Sigma$}}

\def\bOmega{\mbox{\boldmath $\Omega$}}

\def\mbz{\bar{\bz}}
\def\my{\bar{y}}
\def\mx{\bar{x}}

\def\det{\mbox{det}}

\newtheorem{Prop}{Proposition}

\newtheorem{Cor}[Prop]{Corollary}
\newtheorem{Ex}{Example}

\begin{document}

\title{Local Statistical Modeling via Cluster-Weighted Approach with Elliptical Distributions}
\author{Salvatore Ingrassia \and Simona C. Minotti \and Giorgio Vittadini}
\institute{Salvatore Ingrassia \at Dipartimento di Impresa, Culture e Societ\`{a} \\ Universit\`{a} di Catania \\
Corso Italia 55, - Catania (Italy). \email{s.ingrassia@unict.it}
\and Simona C. Minotti \at Dipartimento di Statistica 	\\  Universit\`{a} di Milano-Bicocca \\
Via Bicocca degli
Arcimboldi 8 - 20126 Milano (Italy). \email{simona.minotti@unimib.it}
\and Giorgio Vittadini \at Dipartimento di Metodi Quantitativi per l'Economia e le Scienze Aziendali \\  Universit\`{a} di Milano-Bicocca \\
Via Bicocca degli
Arcimboldi 8 - 20126 Milano (Italy). \email{giorgio.vittadini@unimib.it}
}
\date{Received: date / Accepted: date}

\maketitle

\begin{abstract}
Cluster Weighted Modeling (CWM) is a mixture approach regarding the modelisation of the joint probability of data coming from a heterogeneous population. Under Gaussian assumptions, we investigate statistical properties of CWM from both the theoretical and numerical point of view; in particular, we show that CWM includes as special cases mixtures of distributions and mixtures of regressions. Further, we introduce CWM based on Student-{\em t} distributions providing more robust fitting for groups of observations with longer than normal tails or atypical observations. 
Theoretical results are illustrated using some empirical studies, considering both real and simulated data. 
\end{abstract}

\keywords{Cluster-Weighted Modeling, Mixture Models, Model-Based Clustering.}

\section{Introduction}

Finite mixture models provide a flexible approach to statistical modeling of a wide variety of random phenomena characterized by unobserved heterogeneity. In those models, dating back to the work of \cite{New:1886} and \cite{Pea:1894}, it is assumed that the observations in a sample arise from unobserved groups in the population and the purpose is to identify the groups and estimate the parameters of the conditional-group density functions. If there are no exogenous variables explaining the means and the variances of each component, we refer to unconditional mixture models, i.e. the so-called \textit{Finite Mixtures of Distributions} (FMD), developed for both normal and non-normal components, see e.g. \citet{ EvHa:81, TiSmMa:85, McLaBa:88, McLaPe:00, Schn:05}. Otherwise, we refer to conditional mixture models, i.e. \emph{Finite Mixtures of Regression models} (FMR) and \textit{Finite Mixture of Generalized Linear Models} (FMGLM), see e.g. \citet{DeSCr:88, Jan:93, WeDeSa:95, McLaPe:00, Schn:05}.  These models are also known as mixture-of-experts models in machine learning  \citep{JoJa:94, PeJaTa:96, NgMcLa:07, NgMcLa:08}, switching regression models in econometrics \citep{Quan:72}, latent class regression models in marketing \citep{DeSCron:88, WeKa:00}, mixed models in biology \citep{WPCL:96}. An extension of FMR are the so-called \textit{Finite Mixtures of Regression models with Concomitant variables} (FMRC) \citep{DaMa:88, Wedel:02}, where the weights of the mixture functionally depend on a set of concomitant variables, which may be different from the explanatory variables and are usually modeled by a multinomial logistic distribution. 

The present paper focuses on a different mixture approach, which regards the modelisation of the joint probability of a response variable and a set of explanatory variables. In the original formulation, proposed by \cite{Ger97}, it was called {\em Cluster-Weighted Modeling} (CWM) and was developed in the context of media technology, in order to build a digital violin with traditional inputs and realistic sound \citep{Ger99, Scho00, ScGe01}. \cite{Wedel:02} refers to such model as the {\em saturated mixture regression model}. Moreover, \cite{WeDeSa:02} propose an extensive testing of nested models for market segment derivation and profiling. In the literature, CWM has been developed essentially under Gaussian assumptions; here we set such models in a quite wide framework by considering both direct and indirect applications.

In this paper we reformulate CWM from a statistical point of view and first, we prove that, under suitable assumptions, FMD,  FMR and FMRC are special cases of CWM. Secondly, we introduce CWM based on Student-{\em t} distributions, which have been proposed in the literature in order to provide more robust fitting for groups of observations with longer than normal tails or atypical observations \citep{LLT:89, BeGi:92, McLaPe:00, PeMcLa:00, NaKo:05}.
Theoretical results will be illustrated using  some numerical studies based on both real and simulated data. 

The rest of the paper is organized as follows. In Section \ref{sec:CWMintro} CWM is introduced in a statistical framework; this formulation enables, in Section \ref{sec:CWMGG}, a comparison with FMD, FMR and FMRC under Gaussian assumptions; in Section \ref {sec:CWM-t} Student-$t$ CWM  is introduced and the links with \textit{Finite Mixtures of Student-t distributions} (FMT) are discussed; then, some empirical studies based on both real and simulated datasets are discussed in Section \ref{sec:simul}. Finally, in Section \ref{sec:concl} we provide some conclusions and remarks for further research. In the Appendix a geometrical analysis of 
the decision surfaces of CWM  is reported.

\section{Cluster-Weighted Modeling}\label{sec:CWMintro}

In the original formulation, {\em Cluster-Weighted Modeling} (CWM) was introduced under Gaussian and linear assumptions \citep{Ger97}; here we present CWM in a quite general setting.
Let $(\bX,Y)$ be a pair of a random vector $\bX$ and a random variable $Y$ defined on $\Omega$ with joint probability distribution $p(\bx,y)$, where $\bX$ is the $d$-dimensional input vector with values in some space $\cX \subseteq \mathbb{R}^d$ and $Y$ is a response variable having values in $\mathcal{Y} \subseteq \mathbb{R}$. Thus $(\bx, y) \in \cX \times \cY \subseteq \mR^{d+1}$. Suppose that $\Omega$ can be partitioned into $G$ disjoint groups, say $\Omega_1, \ldots,\Omega_G$, that is $\Omega= \Omega_1 \cup \cdots \cup \Omega_G$.
CWM decomposes the joint probability $p(\bx,y)$ as follows:
\begin{equation}
p(\bx,y; \btheta)=\sum^{G}_{g=1} p(y|\bx,\Omega_g)\, p(\bx|\Omega_g)\, \pi_g, \label{CWM_base}
\end{equation}
where $p(y|\bx,\Omega_g)$ is the conditional density of the response variable $Y$ given the predictor vector $\bx$ and  $\Omega_g$, $p(\bx|\Omega_g)$ is the probability density of $\bx$ given $\Omega_g$, $\pi_g=p(\Omega_g)$ is the mixing weight of $\Omega_g$, ($\pi_g > 0$ and $\sum^{G}_{g=1} \pi_g =1$), $g=1, \ldots, G$. Vector $\btheta$ denotes the set of all parameters of the model.
Hence, the joint density of $(\bX,Y)$ can be viewed as a mixture of local models $p(y|\bx,\Omega_g)$ weighted (in a broader sense) on both the local densities $p(\bx|\Omega_g)$ and the mixing weights $\pi_g$. 

Generalizing some ideas given in \citet{TiSmMa:85}, pag. 2, we can distinguish three types of application of  (\ref{CWM_base}):
\begin{enumerate}
\item {\em Direct application of type A}. We  assume that each  group $\Omega_g$ is characterized by an input-output relation which can be written as $Y|\bx = \mu(\bx; \bbeta_g) + \varepsilon_g$, where $\varepsilon_g$ is a random variable with zero mean and finite variance $\sigma^2_g$, and $\bbeta_g$ denotes the set of parameters of $\mu(\cdot)$ function, $g=1, \ldots, G$. 
\item {\em Direct application of type B}. We assume that there is a random vector $\bZ$ defined on $\Omega= \Omega_1 \cup \cdots \cup \Omega_G$  with values in $\mR^{d+1}$ and each $\bz \in \cZ \subseteq \mR^{d+1}$ belongs to one of these groups. Further, vector $\bz$ is partitioned as  $\bz =(\bx',y)'$ and we assume that within each group we write $p(\bz; \Omega_g) = p((\bx',y)';\Omega_g) = p(y|\bx,\Omega_g)\, p(\bx|\Omega_g)$. In other words,  (\ref{CWM_base}) is another form of density of FMD given by:
\begin{equation}
p(\bz; \btheta) = \sum^{G}_{g=1} p(\bz| \Omega_g) \pi_g = \sum^{G}_{g=1} p(y|\bx,\Omega_g)\, p(\bx|\Omega_g)\, \pi_g. \label{CWM_bmixt}
\end{equation}
\item {\em Indirect application}. In this case,  the density function of CWM \eqref{CWM_base} is simply used as a mathematical tool for density estimation. 
\end{enumerate}
Throughout this paper we concentrate on direct applications. In this case, the posterior probability $p(\Omega_g|\bx,y)$ of unit $(\bx,y)$ belonging to the $g$-th group $(g=1,...,G)$ is given by:
\begin{equation}
 p(\Omega_g|\bx,y) =\frac{p(\bx,y,\Omega_g)}{p(\bx,y)}=\frac{p(y|\bx,\Omega_g)p(\bx|\Omega_g)\pi_g}{\sum^{G}_{j=1}p(y|\bx,\Omega_j)p(\bx|\Omega_j)\pi_j}, \qquad g=1, \ldots, G 
\label{post_base}
\end{equation}
that is the classification of each unit depends on both marginal and conditional densities.
Since $p(\bx|\Omega_g)\pi_g = p(\Omega_g|\bx) p(\bx)$, from (\ref{post_base}) we get:
\begin{align}
 p(\Omega_g|\bx,y) & =\frac{p(y|\bx,\Omega_g)p(\Omega_g|\bx)p(\bx)}{\sum^{G}_{j=1}p(y|\bx,\Omega_j)p(\Omega_j|\bx)p(\bx)}
= \frac{p(y|\bx,\Omega_g)p(\Omega_g|\bx)}{\sum^G_{j=1} p(y|\bx,\Omega_j) p(\Omega_j|\bx)} \: , 
\label{post_scomp}
\end{align}
with
\begin{equation*} 
p(\Omega_g|\bx)   = \frac{p(\bx|\Omega_g) \pi_g}{\sum_{j=1}^G p(\bx|\Omega_j) \pi_j} = \frac{p(\bx|\Omega_g) \pi_g}{p(\bx)} 
\: , \label{post_x}
\end{equation*}
where we set $p(\bx) = \sum_{j=1}^G p(\bx|\Omega_j) \pi_j$.

\subsection{The basic model: linear Gaussian CWM}
In the traditional framework, both marginal densities and conditional densities are assumed to be Gaussian, with
$\bX|\Omega \sim N_d(\bmu_g, \bSigma_g)$ and $Y|\bx, \Omega_g \sim N(\mu(\bx, \bbeta_g), \sigma^2_{\varepsilon, g})$, 
so that we shall write $p(\bx|\Omega_g)=\phi_d(\bx; \bmu_g, \bSigma_g)$  and $p(y|\bx,\Omega_g)=\phi(y; \mu(\bx;\bbeta_g),  \sigma^2_{\varepsilon, g})$, $g=1, \ldots, G$, where the conditional densities are based on linear mappings, so that  $\mu(\bx;\bbeta_g) = \bb'_g \bx + b_{g0}$, for some $\bbeta=(\bb_g',b_{g0})'$, with $\bb \in \mR^d$ and $b_{g0} \in \mR$ . Thus, we get: 
\begin{equation}
p(\bx,y; \btheta)=\sum^{G}_{g=1} \phi(y; \bb_g'\bx+b_{g0},  \sigma^2_{\varepsilon, g}) 
\, \phi_d(\bx; \bmu_g, \bSigma_g)\, \pi_g \: , \label{CWM-GGlin}
\end{equation}
with $\phi(\cdot)$ denoting the probability density of Gaussian distributions.
Model (\ref{CWM-GGlin}) will be referred to as {\em linear Gaussian} CWM.
In particular, the posterior probability  (\ref{post_base}) specializes as:
\begin{equation}
p(\Omega_g|\bx,y)  =\frac{\phi(y; \bb_g'\bx+b_{g0},  \sigma^2_{\varepsilon, g})\phi_d(\bx; \bmu_g, \bSigma_g)\pi_g}{\sum^{G}_{j=1}\phi(y; \bb_j'\bx+b_{j0},  \sigma^2_{\varepsilon, j})\phi_d(\bx; \bmu_j, \bSigma_j)\pi_j} 
\qquad g=1, \ldots, G \: . \label{postCWMGlin}
\end{equation}


\section{Linear Gaussian CWM and relationships with traditional mixture models}\label{sec:CWMGG}

In this section we investigate some relationships between \textit{linear Gaussian} CWM and some Gaussian-based mixture models.
We shall prove that, under suitable assumptions, \textit{linear Gaussian} CWM in (\ref{CWM-GGlin}) leads to the same 
posterior probability of such mixture models. In this sense, we shall say that CWM contains \textit{Finite Mixtures of Gaussians} (FMG), \textit{Finite Mixtures of Regression models} (FMR) and \textit{Finite Mixtures of Regression models with Concomitant variables} (FMRC).

\subsection{\textbf{Finite Mixtures of Distributions}}
Let $\bZ$ be a random vector  defined on $\Omega= \Omega_1 \cup \cdots \cup \Omega_G$ with joint probability distribution $p(\bz)$, where $\bZ$ assumes values in some space $\cZ \subseteq \mathbb{R}^{d+1}$. Assume that density $p(\bz)$ of $\bZ$ has the form of a \textit{Finite Mixture of Distribution}, i.e.  $p(\bz) =\sum^{G}_{g=1} p(\bz|\Omega_g)\pi_g$ where $p(\bz|\Omega_g)$ is the probability density of $\bZ|\Omega_g$ and $\pi_g=p(\Omega_g)$ is the mixing weight of group $\Omega_g$, $g=1, \ldots, G$, see e.g. \citet{McLaPe:00}. Finally, denote with $\bmu_g^{(\bz)}$ and $\bSigma_g^{(\bz)}$ the mean vector and the covariance matrix of $\bZ|\Omega_g$ respectively. 

Now let us set $\bZ=(\bX',Y)'$, where $\bX$ is a random vector with values in $\mR^d$ and $Y$ is a random variable.  Thus, we can write
\begin{equation}
 \bmu_g^{(\bz)} = \begin{pmatrix} \bmu_g^{(\bx)} \\ \mu_g^{(y)} \end{pmatrix} \quad {\rm and } \quad \bSigma_g^{(\bz)} = \begin{pmatrix} \bSigma^{(\bx \bx)}_g & \bSigma^{(\bx \, y)}_g \\
\bSigma^{(y \, \bx)}_g & \sigma^{2(y)}_g \end{pmatrix} \: . \label{Wdecomp}
\end{equation}
Further, the posterior probability is given by:
\begin{equation}
p(\Omega_g|\bz)=\frac{p(\bz|\Omega_g)\pi_g}{\sum^{G}_{g=1}p(\bz|\Omega_g)\pi_g}  
\qquad g=1, \ldots, G \: . \label{postFMG}
\end{equation}
We have the following result:
\begin{Prop}\label{prop:CWM-Z}{\rm Let $\bZ$ be a  random vector defined on $\Omega=\Omega_1 \cup \cdots \Omega_G$ with values in $\mathbb{R}^{d+1}$, and assume that 
$\bZ|\Omega_g \sim N_{d+1} (\bmu_g, \bSigma_g)$ ($g=1, \ldots, G$). In particular, the density $p(\bz)$ of $\bZ$ is a \textit{Finite Mixture of Gaussians} (FMG): 
\begin{equation}
p(\bz) =\sum^{G}_{g=1} \phi_{d+1}(\bz; \bmu_g, \bSigma_g)\pi_g \:. \label{FMG}
\end{equation}
Then, $p(\bz)$  can be written  like \eqref{CWM-GGlin}, that is as a \textit{linear Gaussian} CWM. 
\\ \\{\em Proof.} Let us set $\bZ=(\bX',Y)'$, where $\bX$ is a $d$-dimensional random vector and $Y$ is a random variable.
According to well known results of multivariate statistics, see e.g. \citet{MKB:79}, from  (\ref {FMG}) we get
\begin{align*}
p(\bz)&= \sum^{G}_{g=1} \phi_{d+1}(\bz; \bmu_g, \bSigma_g) \pi_g  = \sum^{G}_{g=1} \phi_{d+1}((\bx,y); \bmu_g, \bSigma_g) \pi_g \notag \\
 & = \sum^{G}_{g=1} \phi_d(\bx; \bmu_g^{(\bx)}), \bSigma_g^{(\bx\bx)}) \phi (y; \mu_g^{(y|\bx)}, \sigma_g^{2(y|\bx)}) \pi_g \: .
\end{align*}
where $\mu_g^{(y|\bx)} = \mu_g^{(y)} + \bSigma_g^{(y\bx)}\bSigma_g^{(\bx \bx)^{-1}}(\bx-\bmu_g^{(\bx)})$ and $\sigma_g^{2(y|\bx)}=\bSigma_g^{(yy)}$. 
If we set $\bb_g=\bSigma_g^{(y\bx)}\bSigma_g^{(\bx \bx)^{-1}}$, $b_{g0}=\mu_g^{(y)} - \bSigma_g^{(y\bx)}\bSigma_g^{(\bx \bx)^{-1}}\bmu_g^{(\bx)}$ and $\sigma^2_{\varepsilon, g} = \sigma_g^{2(y|\bx)}$, then (\ref{FMG}) can be written in the form (\ref{CWM-GGlin}).
$\qed$
}\end{Prop}
Using similar arguments, it can be shown that FMG and \textit{linear Gaussian} CWM lead to the same distribution of posterior probabilities and thus CWM contains FMG. 

\begin{figure}
	\begin{center}\hspace{-1 cm}
	\begin{tabular}{cc}
	 \includegraphics[height=8 cm, width=8 cm]{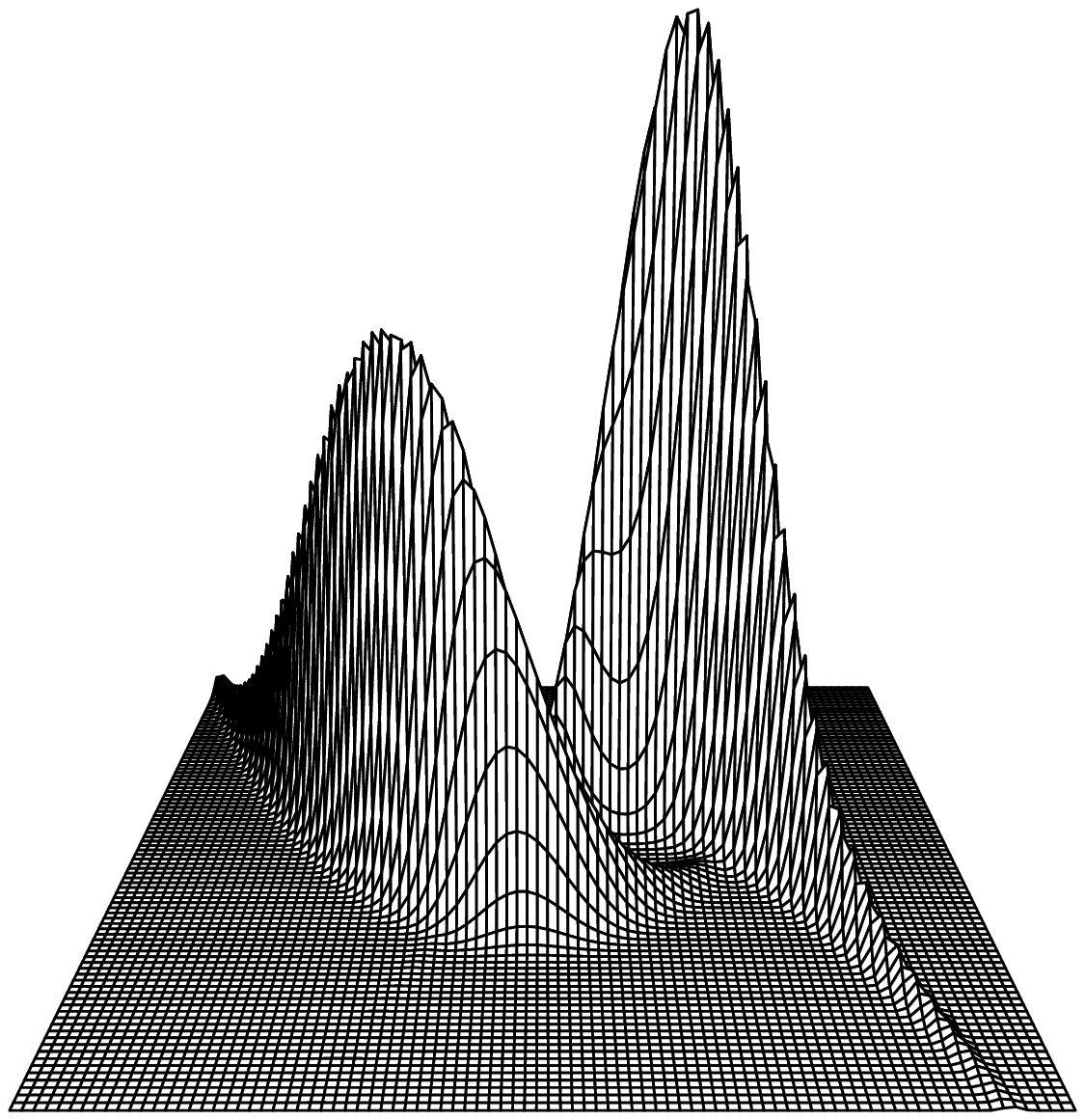} & \includegraphics[height=8 cm, width=8 cm]{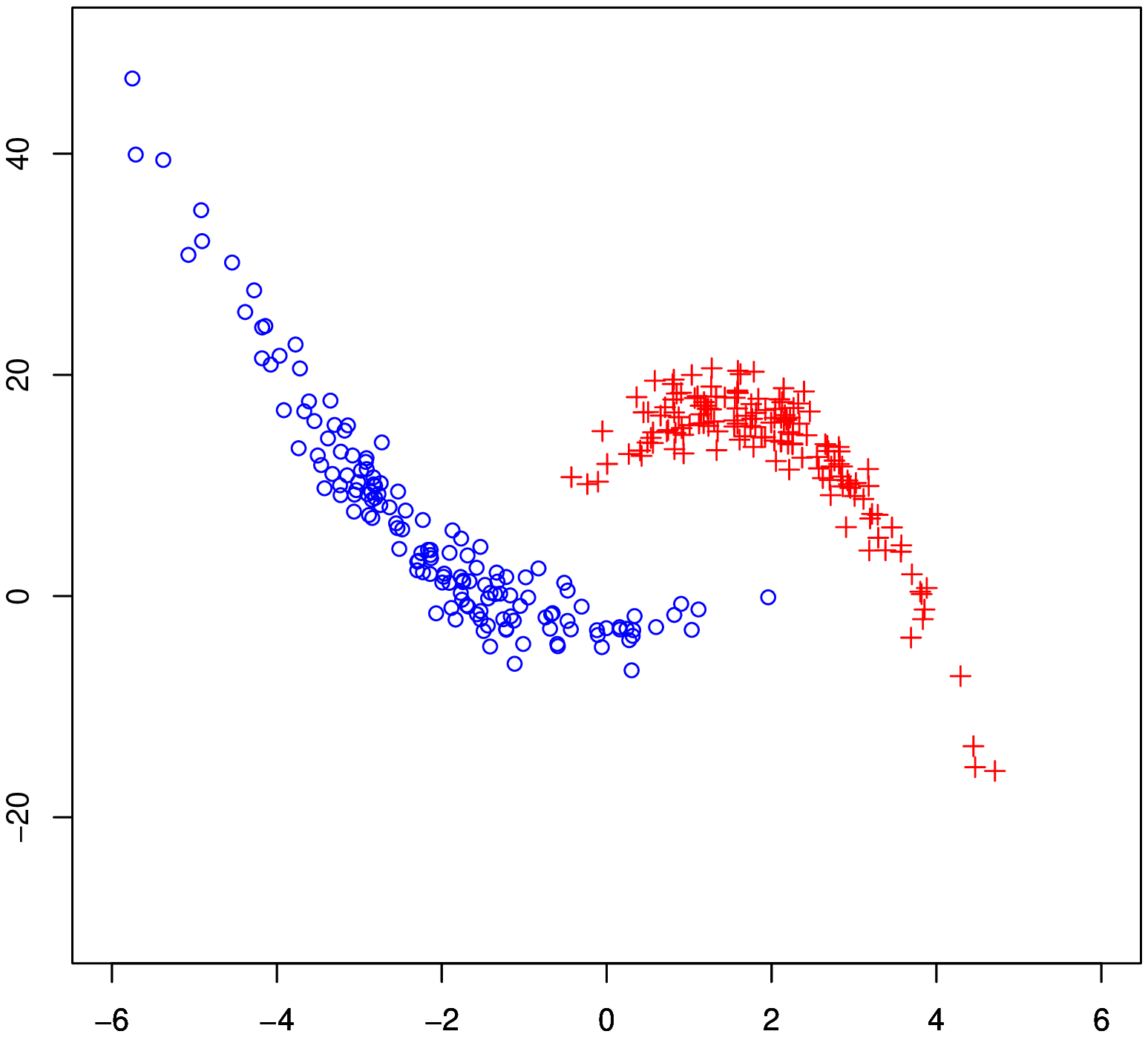} \\
		 \includegraphics[height=8 cm, width=8 cm]{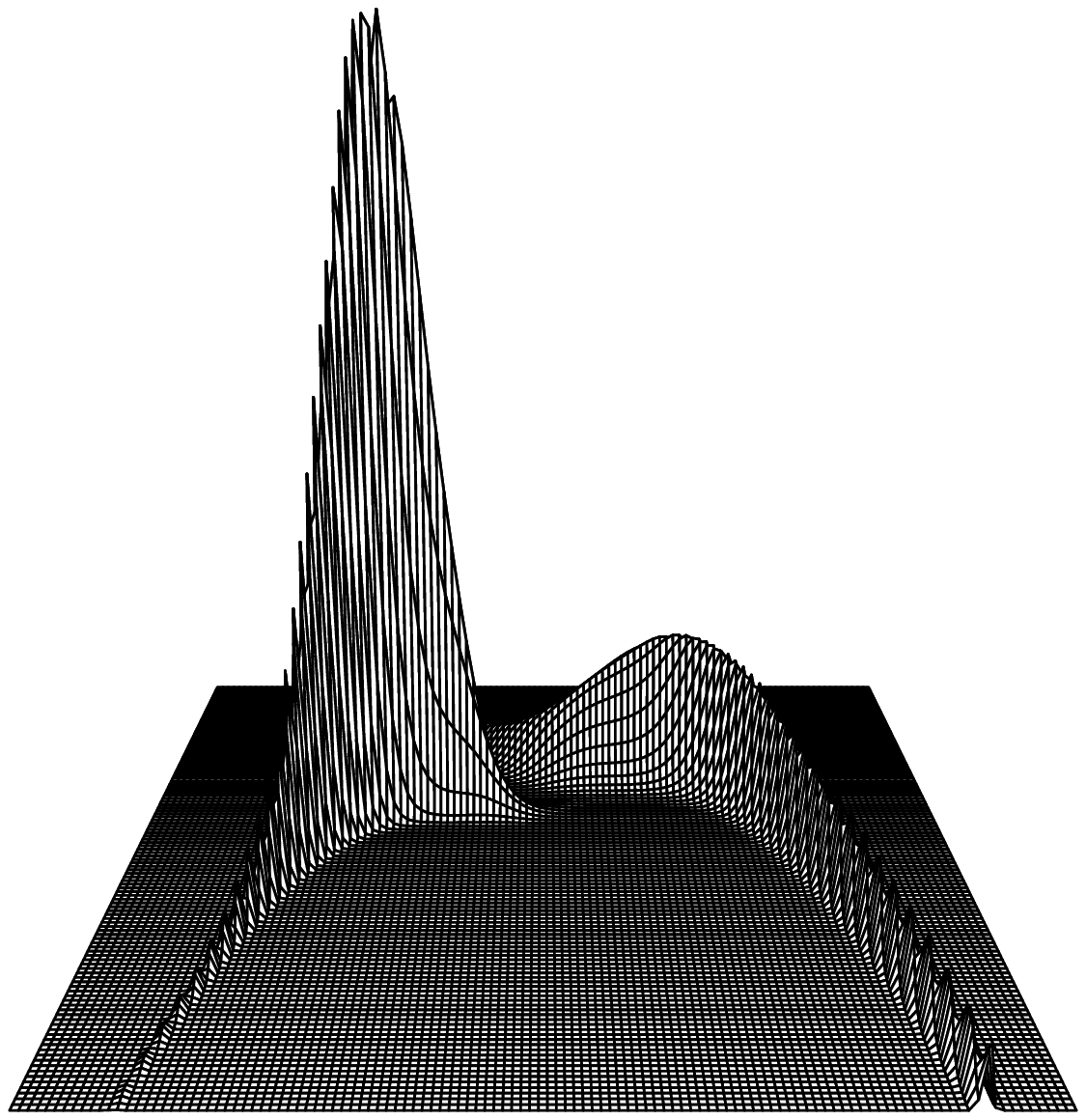} & \includegraphics[height=8 cm, width=8 cm]{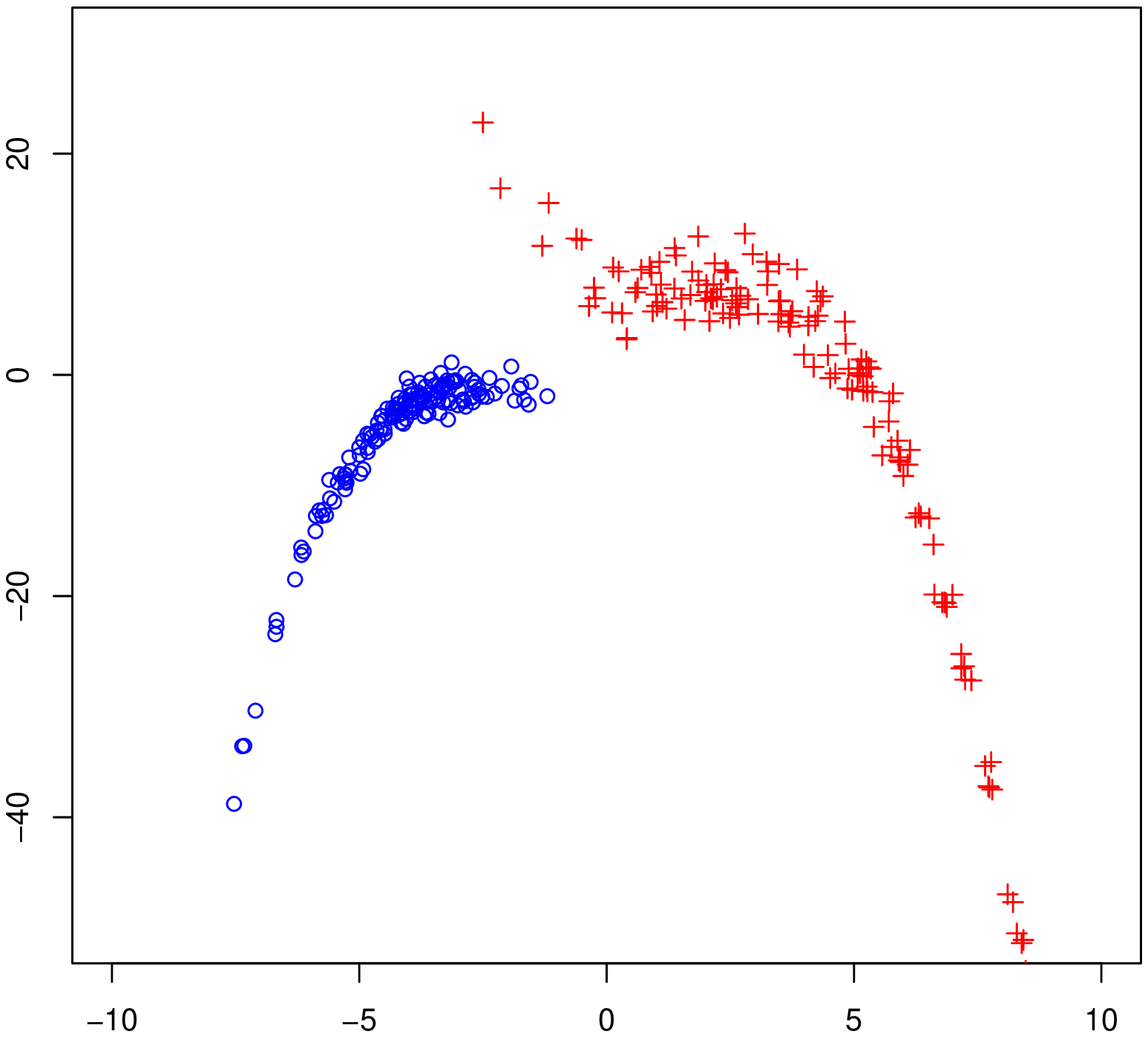} 
	\end{tabular}
	\end{center}
  \caption{\rm Two examples of Gaussian CWM densities and sample data based on quadratic (above) and cubic (below) mappings.}\label{fig:GWMG23}
\end{figure}

We remark that the equivalence between FMG and CWM  holds only for linear mappings  $\mu(\bx;\bbeta_g) = \bb'_g \bx + b_{g0}$ ($g=1, \ldots, G$), while, generally, Gaussian CWM 
\begin{equation}
p(\bx,y; \btheta)=\sum^{G}_{g=1} \phi(y; \mu(\bx;\bbeta_g) ,  \sigma^2_{\varepsilon, g}) 
\, \phi_d(\bx; \bmu_g, \bSigma_g)\, \pi_g \:  \label{CWM-GGgen}
\end{equation}
includes  a quite wide family of FMD. In Figure \ref{fig:GWMG23} we plot two examples of density (\ref{CWM-GGgen}) for both quadratic and cubic functions $\mu(\bx;\bbeta_g)$ (with $G=2$ groups).

\subsection{\textbf{Finite Mixtures of Regression models}}\label{sec:CWM-FMR}
Secondly, {\em Finite Mixtures of regression models} (FMR) \citep{DeSCr:88, McLaPe:00, Schn:05} are considered:
\begin{equation}
f(y|\bx;  \bpsi) = \sum^{G}_{g=1} \phi(y; \bb_g'\bx+b_{g0},  \sigma^2_{\varepsilon, g})\,  \pi_g,  \label{FMR}
\end{equation}
where vector $\bpsi$ denotes the overall parameters of the model. 
The posterior probability $p(\Omega_g|\bx,y)$ of the $g$-th group $(g=1,...,G)$ for FMR is:
\begin{equation}
p(\Omega_g|\bx,y) = \frac{f(y|\bx;  \bpsi,\Omega_g)}{f(y|\bx;  \bpsi)}=\frac{\phi(y; \bb_g'\bx+b_{g0},  \sigma^2_{\varepsilon, g})\pi_g}{\sum^{G}_{j=1}\phi(y; \bb_j'\bx+b_{j0},  \sigma^2_{\varepsilon, j})\pi_j},  
\qquad g=1, \ldots, G \:  \label{postFMR}
\end{equation}
that is the classification of each observation depends on the local model and the mixing weight. We have the following result:
%
\begin{Prop}\label{prop:CWM-FMR}{\rm Let us consider the linear Gaussian CWM (\ref{CWM-GGlin}), with  $\bX|\Omega_g \sim N_d(\bmu_g, \bSigma_g)$ for $g=1,...,G$.
If the probability density of $\bX|\Omega_g$ does not depend on group $g$, i.e. $\phi_d(\bx;\bmu_g,\bSigma_g)= \phi_d(\bx;\bmu,\bSigma)$ for every $g=1,\ldots,G$,   then it follows 
\begin{equation*}
p(\bx,y; \btheta) = \phi_d(\bx;\bmu,\bSigma)  f(y|\bx; \bpsi) \; ,
\end{equation*}
where $f(y|\bx; \bpsi)$ is the FMR model (\ref{FMR}). 
\\ \\{\em Proof.} Assume that $\phi_d(\bx;\bmu_g,\bSigma_g)= \phi_d(\bx;\bmu,\bSigma)$, $g=1, \ldots, G$, then  
(\ref{CWM-GGlin}) yields:
\begin{align*}
p(\bx,y; \btheta) &= \sum^{G}_{g=1} \phi(\bb_g'\bx+b_{g0},  \sigma^2_{\varepsilon, g}) \, \phi_d(\bx; \bmu, \bSigma)\, \pi_g  \notag \\ &= \phi_d(\bx; \bmu, \bSigma)\, \sum^{G}_{g=1} \phi(y; \bb_g'\bx+b_{g0},  \sigma^2_{\varepsilon, g})\, \pi_g  \\
&= \phi_d(\bx; \bmu, \bSigma)\, \sum^{G}_{g=1} f(y|\bx;  \bpsi) \pi_g,
\end{align*}
where $f(y|\bx; \bpsi)$ is FMR model (\ref{FMR}). $\qed$
} \end{Prop}

The second result of this section shows that, under the same hypothesis,  CWM contains FMR. 
\begin{Cor}{\rm 
If the probability density of $\bX|\Omega_g \sim N_d(\bmu_g, \bSigma_g)$ in  (\ref{CWM-GGlin}) does not depend on the group $g$, 
i.e. $\phi_d(\bx;\bmu_g,\bSigma_g)= \phi_d(\bx;\bmu,\bSigma)$ for every $g=1,\ldots,G$, then the posterior probability (\ref{postCWMGlin}) coincides with (\ref{postFMR}). 
\\ \\{\em Proof.} Assume that $\phi_d(\bx;\bmu_g,\bSigma_g)= \phi_d(\bx;\bmu,\bSigma)$, $g=1, \ldots, G$, thus from (\ref{postCWMGlin}) we get
\begin{align*}
p(\Omega_g|\bx,y) &=\frac{\phi(y; \bb_g'\bx+b_{g0},  \sigma^2_{\varepsilon, g})\phi_d(\bx;\bmu,\bSigma) \pi_g}{\sum^{G}_{j=1}\phi(y; \bb_j'\bx+b_{j0},  \sigma^2_{\varepsilon, j}) \phi_d(\bx;\bmu,\bSigma) \pi_j}=\\
&=\frac{\phi(y; \bb_g'\bx+b_{g0},  \sigma^2_{\varepsilon, g}) \phi_d(\bx;\bmu,\bSigma) \pi_g}{\phi_d(\bx;\bmu,\bSigma) \sum^{G}_{j=1}\phi(y; \bb_j'\bx+b_{j0},  \sigma^2_{\varepsilon, j}) \pi_j}=\\
&=\frac{\phi(y; \bb_g'\bx+b_{g0},  \sigma^2_{\varepsilon, g}) \pi_g}{\sum^{G}_{j=1}\phi(y; \bb_j'\bx+b_{j0},  \sigma^2_{\varepsilon, j}) \pi_j} 
\end{align*}
for $g=1, \ldots, G$, which coincides with (\ref{postFMR}). $\qed$
} \end{Cor} 

\subsection{\textbf{Finite Mixtures of Regression models with Concomitant variables}}
{\em Finite Mixtures of Regression models with Concomitant variables} (FMRC) \citep{DaMa:88, Wedel:02} are an extension of FMR:
\begin{equation}
f^*(y|\bx;  \bpsi^*)  = \sum^{G}_{g=1} \phi(y; \bb_g'\bx+b_{g0},  \sigma^2_{\varepsilon, g})\,  p(\Omega_g|\bx, \bxi) \: , \label{FMRC}
\end{equation}
where the mixing weight $ p(\Omega_g|\bx, \bxi)$ is a function depending on $\bx$ through some parameters $\bxi$ and $\bpsi^*$ is the augmented set of all parameters of the model.
The probability $p(\Omega_g|\bx, \bxi) $ is usually modeled by a multinomial logistic distribution with the first component as baseline, that is:
\begin{align}
p(\Omega_g|\bx, \bxi) & = \frac{\exp(\bw_g'\bx+w_{g0})}{\sum^{G}_{j=1} \exp(\bw_j'\bx+w_{j0})}  \, . 
\label{mult_logist}
\end{align}
Equation (\ref{mult_logist}) is satisfied by multivariate Gaussians of $\bX$-distributions with the same covariance matrices, see e.g. \citet{Ande:72}.

The posterior probability $p(\Omega_g|\bx,y)$ of the $g$-th group $(g=1,...,G)$ for FMRC is:
\begin{equation}
p(\Omega_g|\bx,y)  =\frac{f^*(y|\bx;  \bpsi^*,\Omega_g)}{f^*(y|\bx;  \bpsi^*)}=  \frac{\phi(y; \bb_g'\bx+b_{g0},  \sigma^2_{\varepsilon, g}),\sigma_{\epsilon,g}^2) p(\Omega_g|\bx, \bxi)}{\sum_{j=1}^G\phi(y; \bb_j'\bx+b_{j0},  \sigma^2_{\varepsilon, j}) p(\Omega_j|\bx, \bxi)}. \label{postFMRC}
\end{equation}

Under suitable assumptions, \textit{linear Gaussian} CWM leads to the same estimates of $\bb_g, b_{g0}$ ($g=1, \ldots, G$) in (\ref{FMRC}).
\begin{Prop}\label{prop:CWM-FMRC}{\rm 
Let us consider the linear Gaussian CWM (\ref{CWM-GGlin}), with  $\bX|\Omega_g \sim N_d(\bmu_g, \bSigma_g)$ $(g=1,...,G)$. 
If $\bSigma_g=\bSigma$ and $\pi_g=\pi=1/G$ for every $g=1,\ldots,G$, then it follows
\begin{equation*}
p(\bx,y; \btheta) = p(\bx)   f^*(y|\bx; \bpsi^*),
\end{equation*}
where $f^*(y|\bx; \bpsi^*)$ is the FMRC model (\ref{FMRC}) based on the multinomial logistic (\ref{mult_logist}) and $p(\bx) = \sum_{g=1}^G p(\bx|\bOmega_g) \pi_g$.
\\ \\{\em Proof.} Assume $\bSigma_g=\bSigma$ and $\pi_g=\pi=1/G$ for every $g=1,\ldots,G$, thus the density (\ref{CWM-GGlin}) yields:
\begin{align*}
p(\bx,y; \btheta) &= \sum^{G}_{g=1} \phi(\bb_g'\bx+b_{g0},  \sigma^2_{\varepsilon, g}) \, \phi_d(\bx; \bmu_g, \bSigma)\, \pi  \\ 
  & = p(\bx) \sum^{G}_{g=1} \phi(\bb_g'\bx+b_{g0},  \sigma^2_{\varepsilon, g}) \frac{\phi_d(\bx; \bmu_g, \bSigma) \pi}{p(\bx)} \\
  & = p(\bx) \sum^{G}_{g=1} \phi(\bb_g'\bx+b_{g0},  \sigma^2_{\varepsilon, g}) \frac{\exp\left[-\frac{1}{2}(\bx-\bmu_g)'\bSigma^{-1}(\bx-\bmu_g)\right] }{\sum_{j=1}^{G}\exp\left[-\frac{1}{2}(\bx-\bmu_j)'\bSigma^{-1}(\bx-\bmu_j)\right] } \, ,
\end{align*}
where
\begin{align}
& \frac{\exp\left[-\frac{1}{2}(\bx-\bmu_g)'\bSigma^{-1}(\bx-\bmu_g)\right] }{\sum_{j=1}^{G}\exp\left[-\frac{1}{2}(\bx-\bmu_j)'\bSigma^{-1}(\bx-\bmu_j)\right] }  \notag \\
& \qquad = \frac{1}{1+\sum_{j\neq g}\exp\left[-\frac{1}{2}(\bx-\bmu_j)'\bSigma^{-1}(\bx-\bmu_j)+\frac{1}{2}(\bx-\bmu_g)'\bSigma^{-1}(\bx-\bmu_g)\right]}  \notag \\
& \qquad =\frac{1}{1+\sum_{j\neq g}\exp\left[(\bmu_j-\bmu_g)' \bSigma^{-1}\bx -\frac{1}{2}(\bmu_j+\bmu_g)'\bSigma^{-1}(\bmu_j -\bmu_g) \right]} \label{expGauss}
\end{align}
and we recognize that (\ref{expGauss}) can be written in form (\ref{mult_logist}) for suitable constants $\bw_g, w_{g0}$ ($g=1, \ldots, G$). This completes the proof. $\qed$
} \end{Prop}
Based on similar arguments, we can immediately prove that, under the same hypotheses, CWM contains FMRC. 
\begin{Cor}\label{cor:CWM-FMRC}{\rm 
Let us consider the linear Gaussian CWM (\ref{CWM-GGlin}). If $\bSigma_g=\bSigma$  and $\pi_g=\pi=1/G$ for every $g=1,\ldots,G$, then 
posterior probability (\ref{postCWMGlin}) coincides with (\ref{postFMRC}).
\\ \\{\em Proof.} 
First, based on (\ref{mult_logist}),  let us rewrite (\ref{postFMRC}) as
\begin{equation}
p(\Omega_g|\bx,y) =  \frac{\phi(y; \bb_g'\bx+b_{g0},  \sigma^2_{\varepsilon, g}) \exp(\bw_g'\bx+w_{g0})}{\sum_{j=1}^G \phi(y; \bb_j'\bx+b_{j0},  \sigma^2_{\varepsilon, j}) \exp(\bw_j'\bx+w_{j0})}. \label{postFMRCb}
\end{equation}
Assume $\bSigma_g=\bSigma$ and $\pi_g=\pi=1/G$ for every $g=1,\ldots,G$, thus from (\ref{post_scomp}) and (\ref{postCWMGlin}) we get
\begin{align*}
p(\Omega_g|\bx,y) &=\frac{\phi(y; \bb_g'\bx+b_{g0},  \sigma^2_{\varepsilon, g})\phi_d(\bx;\bmu_g,\bSigma) }{\sum^{G}_{j=1}\phi(y; \bb_j'\bx+b_{j0},  \sigma^2_{\varepsilon, j}) \phi_d(\bx;\bmu_j,\bSigma) }\end{align*}
and after some algebra we find a quantity like  (\ref{postFMRCb}). This completes the proof. $\qed$
} \end{Cor}


As for the relation between FMRC and \textit{linear Gaussian} CWM, consider that the joint density $p(\bx, \Omega_g)$ can be written in either form:
\begin{equation}
p(\bx, \Omega_g) = p(\bx|\Omega_g)p(\Omega_g) \qquad {\rm or} \qquad  p(\bx, \Omega_g) = p(\Omega_g|\bx)p(\bx) \:,
\end{equation}
where quantity $p(\bx|\Omega_g)$ is involved in CWM (left-hand side), while FMRC contains conditional probability $p(\Omega_g|\bx)$ (right-hand side).
In other words,  CWM is a $\Omega_g$-to-$\bx$ model, while  FMRC is a $\bx$-to-$\Omega_g$ model. According to \cite{Jord:95}, in the framework of neural networks, they are called the {\em generative direction} model and the {\em diagnostic direction} model, respectively.

The results of this section are listed in Table \ref{tab:compare}, which summarizes the relationships between \textit{linear Gaussian} CWM and traditional Gaussian mixture models.
\begin{table}[h!]
\begin{center}
\begin{tabular}{llllll}
\hline\hline
model & $p(\bx|\Omega_g)$ & $p(y|\bx,\Omega_g)$  &parameterisation of $\pi_g$ & relationship &assumptions \\ \hline
FMG & Gaussian & Gaussian  & none  & linear & \\
FMR & none & Gaussian  & none & linear & $(\bmu_g, \bSigma_g)=(\bmu, \bSigma)$, g=1,\ldots,G\\
FMRC & none & Gaussian  & logistic & linear & $\bSigma_g=\bSigma$ and $\pi_g=\pi$, g=1,\ldots,G \\
\hline\hline
\end{tabular}
\end{center}
\caption{Relationships between \textit{linear Gaussian} CWM and traditional Gaussian mixtures.}\label{tab:compare}
\end{table}

Finally, we remark that if the conditional distributions $p(y|\bx, \Omega_g)=\phi(y; \bb_g'\bx+b_{g0},  \sigma^2_{\varepsilon, g})$ ($g=1, \ldots, G$) do not depend on group $g$, that is $\phi(y; \bb_g'\bx+b_{g0},  \sigma^2_{\varepsilon, g})=
\phi(y; \bb'\bx+b_{0},  \sigma^2_{\varepsilon})$ for $g=1, \ldots, G$, then (\ref{CWM-GGlin})  specializes as:
\begin{align}
p(\bx,y; \btheta) &=\sum^{G}_{g=1} \phi(y; \bb_g'\bx+b_{g0},  \sigma^2_{\varepsilon, g}) 
\, \phi_d(\bx; \bmu_g, \bSigma_g)\, \pi_g \notag \\
& = \phi(y; \bb'\bx+b_{0},  \sigma^2_{\varepsilon}) \sum^{G}_{g=1} 
\, \phi_d(\bx; \bmu_g, \bSigma_g)\, \pi_g \: , \label{CWM-GGlin=}
\end{align}
and this implies that, from (\ref{FMR}) and (\ref{FMRC}), FMR and FMRC are reduced to a single straight line:
\begin{equation*}
f(y|\bx;  \bpsi) = f^*(y|\bx;  \bpsi^*) =  \phi(y; \bb'\bx+b_{0}),
\end{equation*}
since $\sum_{g=1}^G \pi_g = \sum_{g=1}^G p(\Omega_g|\bx, \bxi)=1$.
The results of this section will be illustrated from a numerical point of view in Section \ref{sec:simCWMGG}.

\section{Student-$t$-CWM}\label{sec:CWM-t}

In this section we introduce CWM based on Student-$t$ distributions.  Data modeling according to the Student-$t$ distribution  has been proposed in the literature in order to provide more robust fitting for groups of observations with longer than normal tails or atypical observations, see e.g. \citet{Zell:76, LLT:89, BeGi:92, McLaPe:98, McLaPe:00, PeMcLa:00, NaKo:05}. Recent applications include also analysis of orthodontic data via linear effect models \citep{PLW:01}, marketing data analysis \citep{AAC:02} and asset pricing \citep{KaZh:06}. 


To begin with, we recall that a $q$ variate  random vector $\bZ$ has a multivariate $t$ distribution  with degrees  of freedom $\nu \in (0, \infty)$, location parameter $\bmu \in \mR^q$ and $q \times q$ positive definite inner product matrix $\bSigma$ if it has density 
\begin{equation}
p(\bz; \bmu, \bSigma, \nu) = \frac{\Gamma((\nu+q)/2) \nu^{\nu/2}}{\Gamma(\nu/2)|\pi\bSigma|^{1/2}[\nu+
\delta(\bz, \bmu; \bSigma)]^{(\nu+q)/2}}, \label{tq-pdf}
\end{equation}
where $\delta(\bz, \bmu; \bSigma) = (\bz-\bmu)'\bSigma^{-1}(\bz-\bmu)$
denotes the squared Mahalanobis distance between $\bz$ and $\bmu$, with respect to matrix $\bSigma$, and $\Gamma(\cdot)$ is the Gamma function.
In this case we write $\bZ \sim t_q(\bmu, \bSigma, \nu)$, and it results $\EE(\bZ)=\bmu$ (for $\nu>1$) and Cov$(\bZ)=\nu \bSigma/(\nu-2)$ (for $\nu>2$).
It is well known that, if $U$ is a random variable, independent of $\bZ$, such that $\nu U$ has a chi-squared distribution with $\nu$ degrees of freedom, that is $\nu U \sim \chi^2_{\nu}$, then 
$ \bZ|(U=u) \sim N_q(\bmu, \bSigma/u)$.

Throughout this section we assume that  $\bX|\Omega_g$ has a multivariate $t$ distribution with location parameter $\bmu_g$, inner product matrix $\bSigma_g$ and degrees
of freedom $\nu_g$, that is $\bX|\Omega_g  \sim t_d(\bmu_g, \bSigma_g, \nu_g)$,  and that $Y|\bx,\Omega_g$ has a $t$ distribution with
location parameter $\mu(\bx; \bbeta_g)$, scale parameter $\sigma^2_g$ and degrees of freedom $\zeta_g$, that is $Y|\bx,\Omega_g\sim
t(\mu(\bx; \bbeta_g), \sigma^2_g, \zeta_g)$, $g=1, \ldots, G$.
Thus (\ref{CWM_base}) specializes as:
\begin{equation}
p(\bx,y; \btheta)=\sum^{G}_{g=1} t(y; \mu(\bx; \bbeta_g),  \sigma^2_g, \zeta_g) \: 
 t_d(\bx; \bmu_g, \bSigma_g,\nu_g)\, \pi_g \:, \label{CWM-tt}
\end{equation}
and this model will be referred to as {\em t}-CWM.
The special case in which $\mu(\bx; \bbeta_g)$ is a linear mapping will be called {\em linear t}-CWM:
\begin{equation}
p(\bx,y; \btheta)=\sum^{G}_{g=1} t(y; \bb'_g \bx +b_{g0},  \sigma^{2}_g, \zeta_g) \, , 
\, \bt_d(\bx; \bmu_g, \bSigma_g,\nu_g)\, \pi_g \: .  \label{CWM-ttlin}
\end{equation}
where, according to (\ref{tq-pdf}), $g=1, \ldots, G$, we have
\begin{align*}
t(y; \bb'_g \bx +b_{g0},  \sigma^{2}_g, \zeta_g) & = \frac{\Gamma((\nu_{g,y}+1)/2)\zeta_g^{\zeta_g/2}}{ \Gamma(\nu_{g,y}/2)\sqrt{\pi\sigma_{\epsilon,g}^2}\{\zeta_g+[y-(\bb_g'\bx+b_{g0})]^2/\sigma_{\epsilon,g}^2\}^{(\zeta_g+1)/2}} \\
t_d(\bx; \bmu_g, \bSigma_g,\nu_g) & = \frac{\Gamma((\nu_{g,\bx}+q)/2)) \nu_{g,\bx}^{\nu_{g,\bx}/2}}{\Gamma(\nu_{g,\bx}/2)|\pi \bSigma_g|^{1/2}\{\nu_{g}+
\delta(\bx, \bmu_g; \bSigma_g)\}^{(\nu_{g,\bx}+q)/2}}.
\end{align*}
Moreover,  the posterior probability (\ref{post_base}) specializes as:
\begin{equation*}
p(\Omega_g|\bx,y)  =\frac{t(y; \bb'_g \bx +b_{g0},  \sigma^{2}_g, \zeta_g) \bt_d(\bx; \bmu_g, \bSigma_g,\nu_g)\pi_g}{\sum^{G}_{j=1}t(y; \bb'_j \bx +b_{j0},  \sigma^{2}_j, \zeta_j) \bt_d(\bx; \bmu_j, \bSigma_j,\nu_j)\pi_j} \, ,
\qquad g=1, \ldots, G \: .
\end{equation*}

The following result implies that, differently from the Gaussian case, \textit{linear t}-CWM is not a \textit{Finite Mixture of $t$-distributions} (FMT).
\begin{Prop}\label{prop:CWM-t}{\rm Let $\bZ$ be a  random vector defined on $\Omega=\Omega_1 \cup \cdots \Omega_G$ with values in $\mathbb{R}^{d+1}$ and set $\bZ=(\bX',Y)'$ where $\bX$ is a $d$-dimensional input vector and $Y$ is a random variable defined on $\Omega$. Assume that the density of $\bZ=(\bX',Y)'$ can be written in the form of a \textit{linear $t$}-CWM (\ref{CWM-ttlin}), where $\bX|\Omega_g  \sim t_d(\bmu_g, \bSigma_g, \nu_g)$   and  $Y|\bx,\Omega_g\sim t(\mu(\bx; \bbeta_g), \sigma^2_g, \zeta_g)$, $g=1, \ldots, G$.
If $\zeta_g=\nu_g+d$ and $\sigma^{*2}_g=\sigma^2_g[\nu_g+\delta(\bx;\bmu_g, \bSigma_g)]/(\nu_g+d)$ then  the \textit{linear t}-CWM 
 (\ref{CWM-ttlin}) coincides with a FMT for suitable parameters $\bb_g, b_{g0}$ and $\sigma^2_g$,  $g=1, \ldots, G$. 
\\ \\{\em Proof.} Let  $\bZ$ be a $q$-variate random vector having multivariate $t$ distribution (\ref{tq-pdf}) with degrees 
of freedom $\nu \in (0, \infty)$, location parameter $\bmu$ and positive definite inner product matrix $\bSigma$. 
If $\bZ$ is partitioned as $\bZ=(\bZ'_1,\bZ'_2)'$, where $\bZ_1$ takes values in $\mR^{q_1}$ and $\bZ_2$ in $\mR^{q_2}=\mR^{q-q_1}$, then $\bZ$ can be written as
\begin{equation*}
 \bmu = \begin{pmatrix} \bmu_1 \\ \bmu_2 \end{pmatrix} \quad {\rm and } \quad \bSigma = \begin{pmatrix} \bSigma_{11} & \bSigma_{12} \\
\bSigma_{21} & \bSigma_{22} \end{pmatrix} \: . 
\end{equation*}
Hence, based on properties of the multivariate $t$ distribution, see e.g. \cite{Dick:67}, \cite{LiRu:95}, it can be proved that:
\begin{equation}
\bZ_1 \sim t_{q_1}(\bmu_1, \bSigma_{11},\nu) \quad {\rm and} \quad \bZ_2|\bz_1 \sim t_{q_2}(\bmu_{2|1}, \bSigma^*_{2|1}, \nu+q_1), \: 
\label{Tdecomp}
\end{equation}
where
\begin{equation}
\begin{split}
\bmu_{2|1} & = \bmu_{2|1} (\bz_1) =\bmu_2+\bSigma_{21}\bSigma_{11}^{-1}(\bz_1-\bmu_1) \\ 
 \bSigma^*_{2|1} & = \bSigma^*_{2|1} (\bz_1) = \frac{\nu+ \delta(\bz_1; \bmu_1, \bSigma_{11})}{\nu+q_1}\bSigma_{2|1}, \: 
\end{split}
 \label{Tpar2|1}
\end{equation}
with $\bSigma_{2|1}=\bSigma_{22}-\bSigma_{21}\bSigma_{11}^{-1}\bSigma_{12}$ and $\delta(\bz_1; \bmu_1, \bSigma_{11}) =(\bz_1-\bmu_1)'\bSigma^{-1}_{11}(\bz_1-\bmu_1)$.
In particular, if we set $\bZ=(\bX',Y)'$ then  (\ref{CWM-ttlin}) coincides with a FMT if $\zeta_g=\nu_g+d$ and $\sigma^{*2}_g=\sigma^2_g[\nu_g+\delta(\bx;\bmu_g, \bSigma_g)]/(\nu_g+d)$. 
$\qed$
}\end{Prop}
In conclusion, we remark that  (\ref{CWM-ttlin}) defines a wide family of densities which includes FMT and \textit{Finite Mixtures of Regression models with Student-$t$ errors} (not proposed in the literature yet,
as far as the authors know). A numerical analysis concerning a comparison between the classifications obtained by either FMT and linear $t$-CWM will
be presented in Section \ref{sec:t-compare}.

\section{Empirical studies}\label{sec:simul}

The statistical models introduced before have been evaluated on the grounds of many empirical studies based on  both real and simulated datasets. The CWM parameters have been estimated  by means of the EM algorithm according to the maximum likelihood approach; the routines have been implemented in R with different initialization strategies, in order to avoid local optima. 
This section is organized as follows. In Subsection \ref{sec:simCWMGG} we consider a comparison among \textit{linear Gaussian} CWM and FMR, FMRC; in Subsection \ref{sec:t-compare} we consider a comparison between  \textit{linear $t$}-CWM and FMT; in Subsection \ref{sec:noisydata} we present further numerical studies based on simulated data in order to evaluate model performance in the area of robust clustering; finally, in Subsection \ref{sec:realdata} we consider another case study concerning a real dataset in the medical area.

\subsection{A comparison among \textit{linear Gaussian} CWM and FMR, FMRC}\label{sec:simCWMGG}
To begin with, we present some simulation studies in order to verify empirically the theoretical results of Section \ref{sec:CWMGG}, that is \textit{linear Gaussian} CWM contains FMR and FMRC as special cases (in particular, data modeling via FMRC has been carried out by means of Flexmix R-package, see \citet{Leisch:04}).

For this aim, we have considered  some cases concerning classification of units $(x,y) \in \mR^2$. 
The data have been obtained as follows: first, we have generated samples $\bx_1,\ldots,\bx_{G}$ (corresponding to the $X$-variable) according to $G$ Gaussian distributions with parameters $(\mu_g,\sigma^2_g)$; each sample $\bx_g$ has a size $N_g$, $(g=1,...,G)$. Afterwards, from  $\bx_g=(x_{g1}, \ldots, x_{gN_g})'$ we obtained vector $\by_g=(y_{g1}, \ldots, y_{gN_g})'$ considering realizations of the random variables $Y_{gn}= b_{g1} x_{gn} +b_{g0} + \epsilon_{g}$, for $g=1, \ldots, G$ and $n=1, \ldots, N_g$, where $b_{g1}, b_{g0}\in \mR$ and $\epsilon \sim N (0, \sigma^2_{\epsilon,g})$. 
Moreover, the obtained results according to CWM, FMR or FMRC have been compared  by means of the following quantities :
\begin{enumerate}
\item The Wilks's lambda $\Lambda$, which is used in multivariate analysis of variance, given by
\begin{equation} 
\Lambda=\frac{\det \, \bW}{\det \, \bT} \label{lambdaW},
\end{equation} where $\bT$ is the total scatter matrix and $\bW$ is the within-class scatter matrix:
\begin{align*}
 \bW = \sum_{g=1}^G \sum_{n=1}^{N_g} (\bz_{ng} - \mbz_g)(\bz_{ng} - \mbz_g)'  \quad {\rm and} \quad \bT = \sum_{g=1}^G \sum_{n=1}^{N_g} (\bx_{ng} - \mbz)(\bz_{ng} - \mbz)',
\end{align*} 
with $\bz=(\bx,y)$ and $\mbz_g=(\mathbf{\mx}_g, \my_g)$ being the vector mean of the $g$-th group.
\item An {\em  index of weighted model fitting} (IWF) $\cE$ defined as:
\begin{equation}
\cE  = \left(\frac{1}{N}\sum_{n=1}^{N} \left[y_n - \left( \sum_{g=1}^{G} \mu(\bx_n; \bbeta_g) p(\Omega_g|\bx_n, y_n) \right) \right]^2 \right)^{1/2}\: . \label{err2}
\end{equation}
\item The misclassification rate $\eta$, that is the percentage of units being classified in the wrong class.
\end{enumerate}

\begin{Ex}\label{retteV}{\rm
Here we give an example in which  the densities of $\bX|\Omega_g$ are not the same, and we show that CWM outperforms FMR.  For this aim, we
considered a two groups data sample; the data have been generated according to the following parameters:
\begin{center}
\begin{tabular}{ccrcccrc}
\hline\hline
				& & \multicolumn{2}{c}{$\phi(x;\mu_i, \sigma^2_g)$} & \qquad & \multicolumn{3}{c}{$\phi(y; b_{g0}+b_{g1}x,   \sigma^2_{\varepsilon, g})$}  \\ \cline{3-4} \cline{6-8}
				& $N_g$ & $\mu_g$ & $\sigma_g$ & & $b_{g0}$ & $b_{g1}$ & $\sigma_{\epsilon,g}$ \\ \hline
{\em Group 1} & 100 &   10  & 2 & & 2 & 6 & 2 \\ 
{\em Group 2} & 200 & $-10$ & 2 & & 4 & $-6$ & 2 \\
\hline\hline
\end{tabular}
\end{center}
where $\mu_2=-\mu_1$. Data are shown in Figure \ref{fig:retteV}.
The obtained results according to  CWM and FMR are summarized in the following table:
\begin{center}
\begin{tabular}{cccc}
\hline\hline
  & $\Lambda$ & $\cE$ & $\eta$ \\ \hline
CWM &   0.0396 & 2.003 &  0.00\%\\
FMR &   0.2306 & 7.013 & 5.33\% \\
\hline\hline
\end{tabular}
\end{center}
The analysis shows that CWM leads to well separated groups ($\Lambda=0.0396$) where all units have been classified properly ($\eta=0.00\%$) and the local models $p(y|x, \Omega_g)$ present a good fitting to data ($\cE=2.003$). On the contrary, FMR presents a bad data fitting ($\cE=7.013$) and it leads to  groups being partially overlapped ($\Lambda=0.2306$), even if the misclassification rate
is small ($\eta=5.33\%$). Finally, we remark that FMRC leads to the same classification of CWM.
The scatter plots of data classified according to CWM and FMR are given in Figure \ref{fig:retteV}. 
\begin{figure}[h] 
	\begin{center}
	\begin{center}
		\includegraphics[height=6cm, width=6 cm]{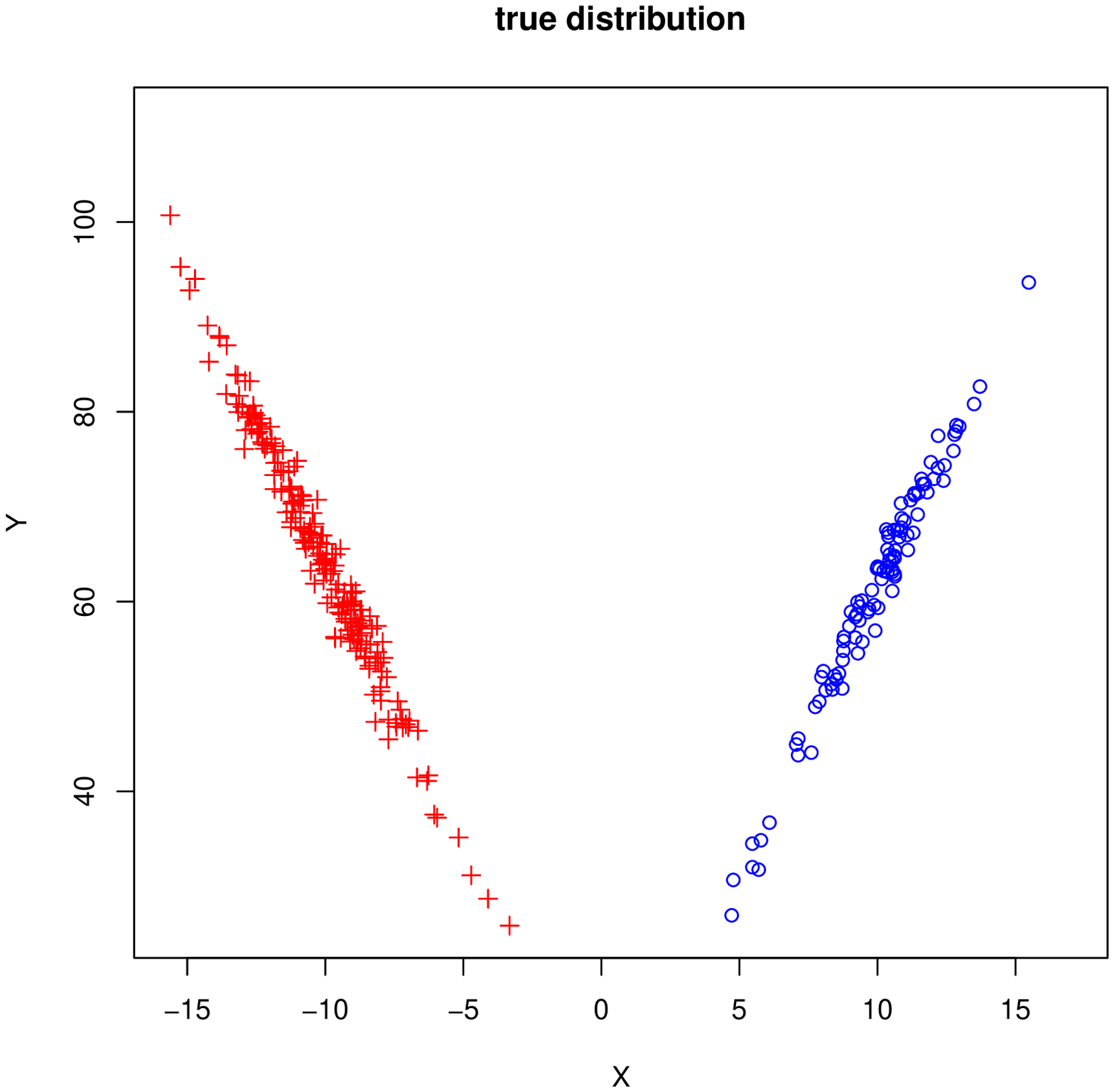} 
	\end{center}
	\begin{tabular}{cc}
		\includegraphics[height=6cm, width=6 cm]{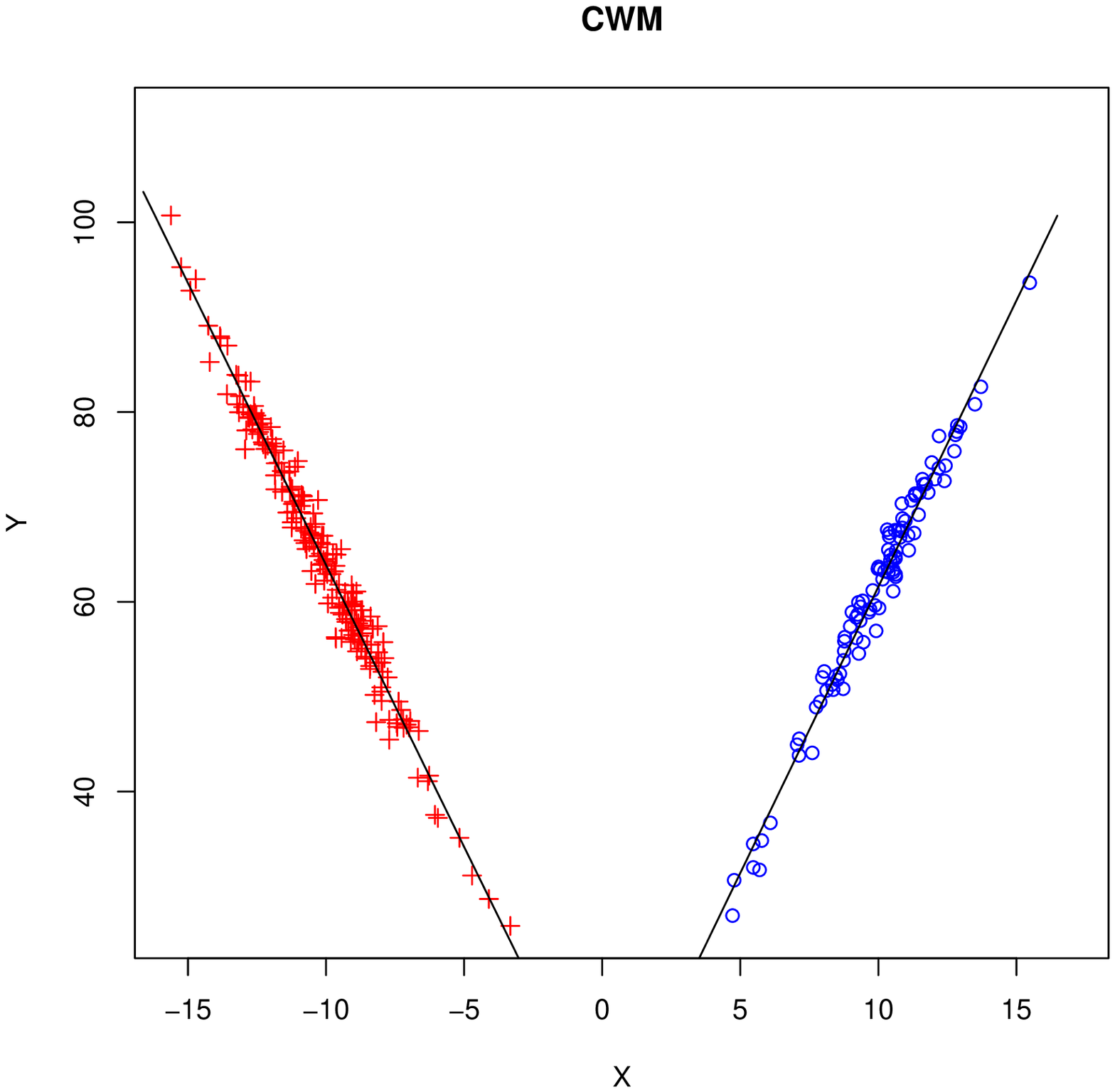} &
		\includegraphics[height=6cm, width=6 cm]{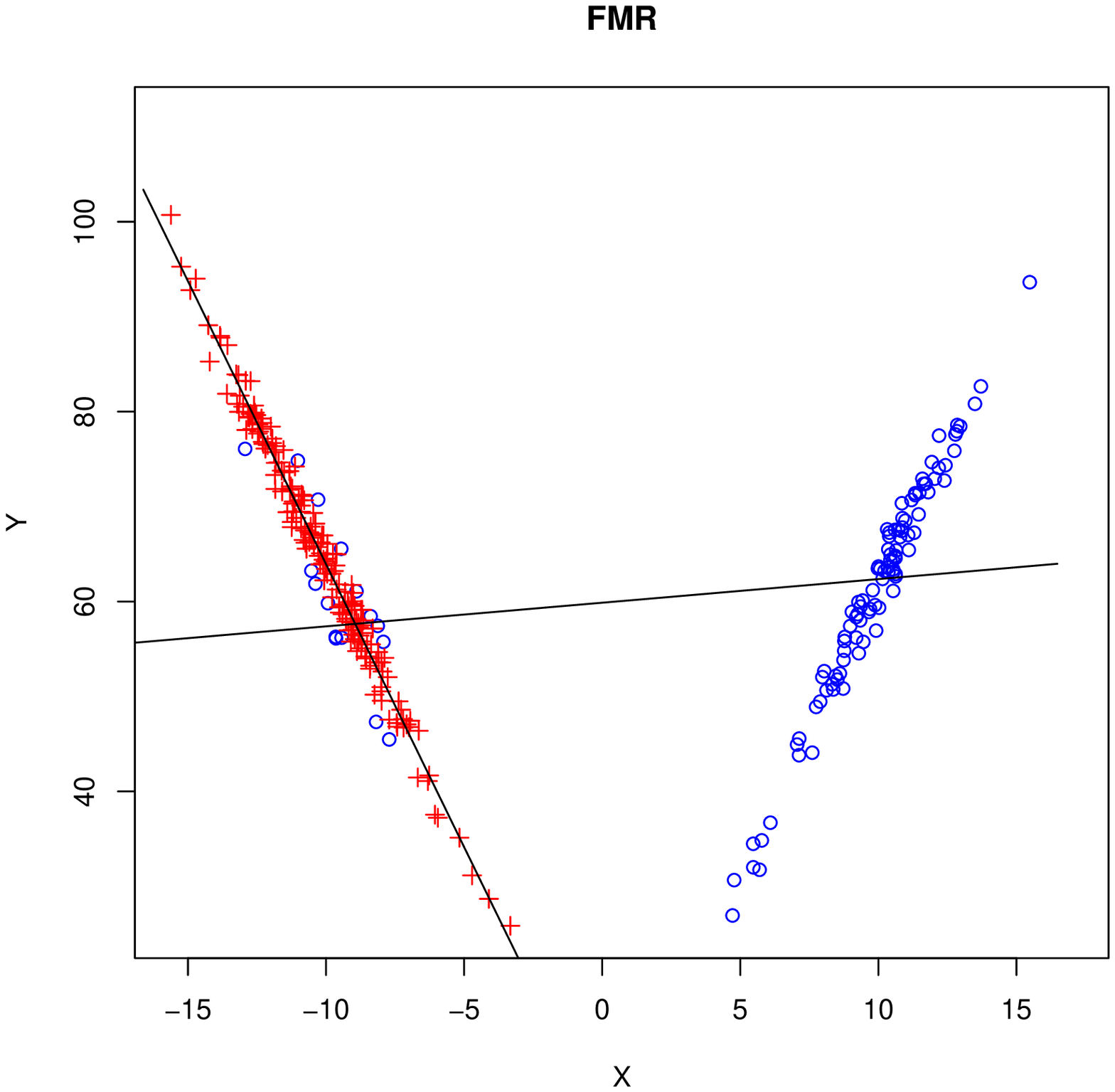} 
		\end{tabular}
	\end{center}
	\vspace{-.5 cm}
  \caption{ Example \ref{retteV}. True distribution, data classification and fitted lines according to CWM and FMR. The symbol $+$ denotes data classified in group 1 and $\circ$ denotes data classified in group 2.}\label{fig:retteV}
\end{figure}

}\end{Ex}

\begin{Ex}\label{rettetirng}{\rm
In order to illustrate a different situation, we consider another example in which the densities of $\bX|\Omega_g$ are not the same. The data have been generated based on the following parameters:
\begin{center}
\begin{tabular}{ccrcccrc}
\hline\hline
				& & \multicolumn{2}{c}{$\phi(x;\mu_i, \sigma^2_g)$} & \qquad & \multicolumn{3}{c}{$\phi(y; b_{g0}+b_{g1}x,   \sigma^2_{\varepsilon, g})$}  \\ \cline{3-4} \cline{6-8}
				& $N_g$ & $\mu_g$ & $\sigma_g$ & & $b_{g0}$ & $b_{g1}$ & $\sigma_{\epsilon,g}$ \\ \hline
{\em Group 1} & 100 &   5  & 1 & & 40 & 6 & 2 \\ 
{\em Group 2} & 200 & $10$ & 2 & & 40 & $-1.5$ & 1 \\
{\em Group 3} & 150 & $20$ & 3 & & 150 & $7$ & 2 \\
\hline\hline
\end{tabular}
\end{center}
Data are shown in Figure \ref{fig:retteV}.
The obtained results according to  CWM and FMR are summarized in the following table:
\begin{center}
\begin{tabular}{cccc}
\hline\hline
  & $\Lambda$ & $\cE$ & $\eta$ \\ \hline
CWM &   0.0498 & 1.678 &  0.00\%\\
FMR &   0.0909 & 1.647 & 8.67\% \\
\hline\hline
\end{tabular}
\end{center}
In this case, the analysis shows that CWM leads to well separated groups ($\Lambda=0.0498$) where all units have been classified properly ($\eta=0.00\%$) and the local models $p(y|x, \Omega_g)$ present a good fitting to data  ($\cE=1.678$). Also FMR  yields essentially the same conditional distributions ($\cE=1.647$), but it presents a slightly worse
value of the $\Lambda=0.0909$ and a  misclassification rate $\eta=8.67\%$. Also in this case, we remark that FMRC attains the same classification of CWM. 
The scatter plots of data classified according to CWM and FMR  are given 
in Figure \ref{fig:rettetriang}. 
\begin{figure}[h] 
	\begin{center}
		\includegraphics[height=6cm, width=6 cm]{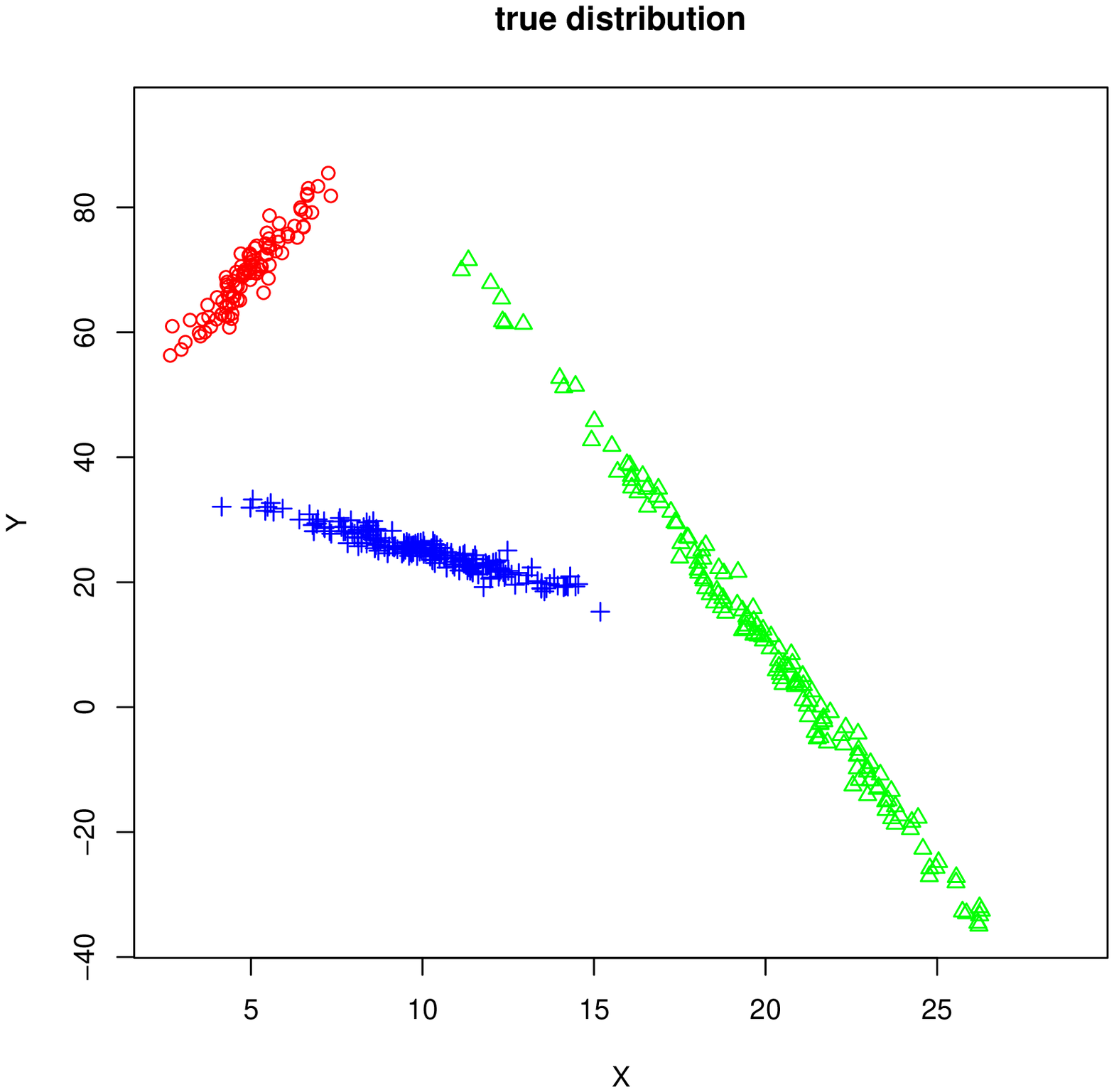} 
	\end{center}
	\begin{center}
		\includegraphics[height=6cm, width=6 cm]{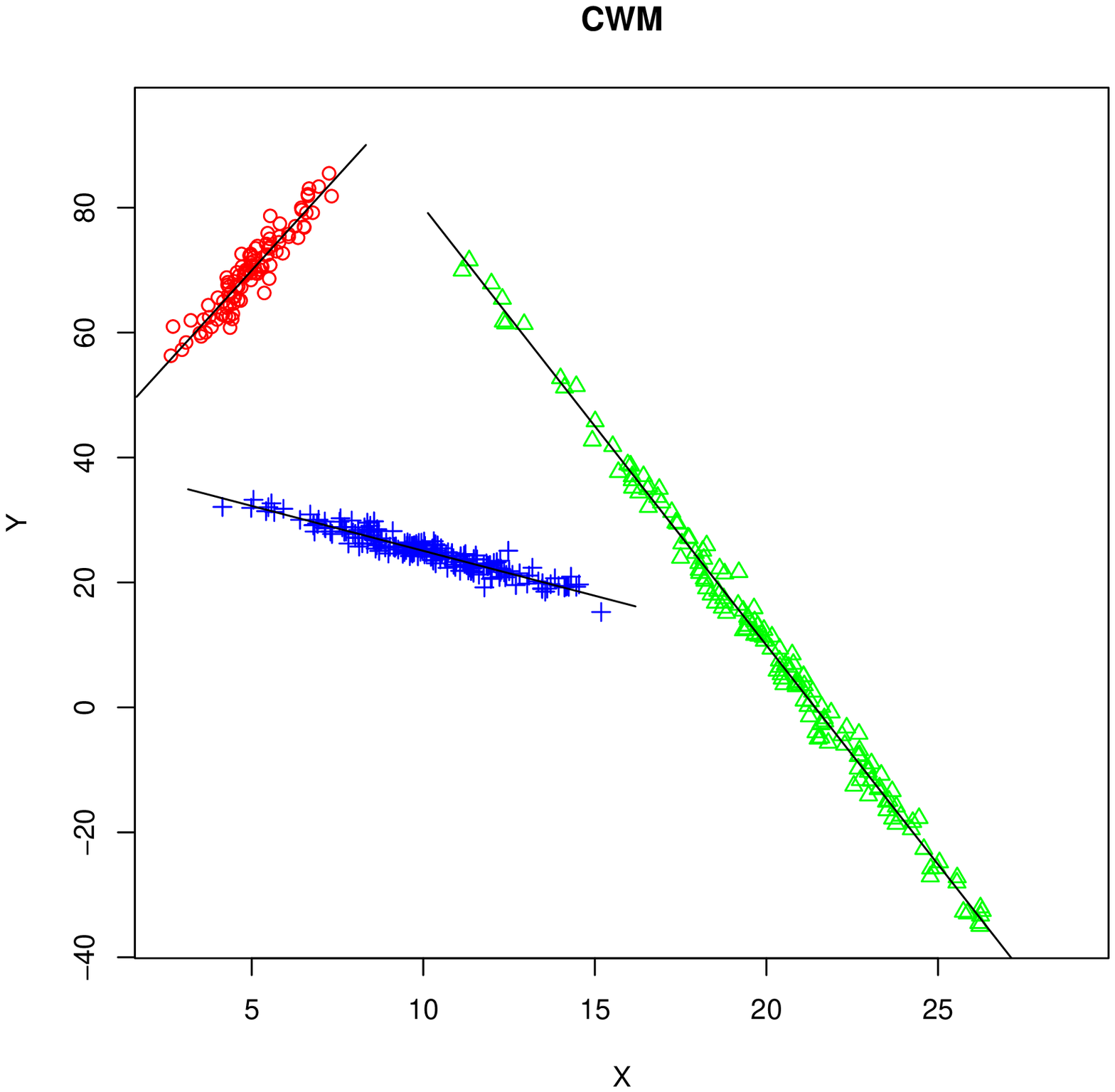}
		\includegraphics[height=6cm, width=6 cm]{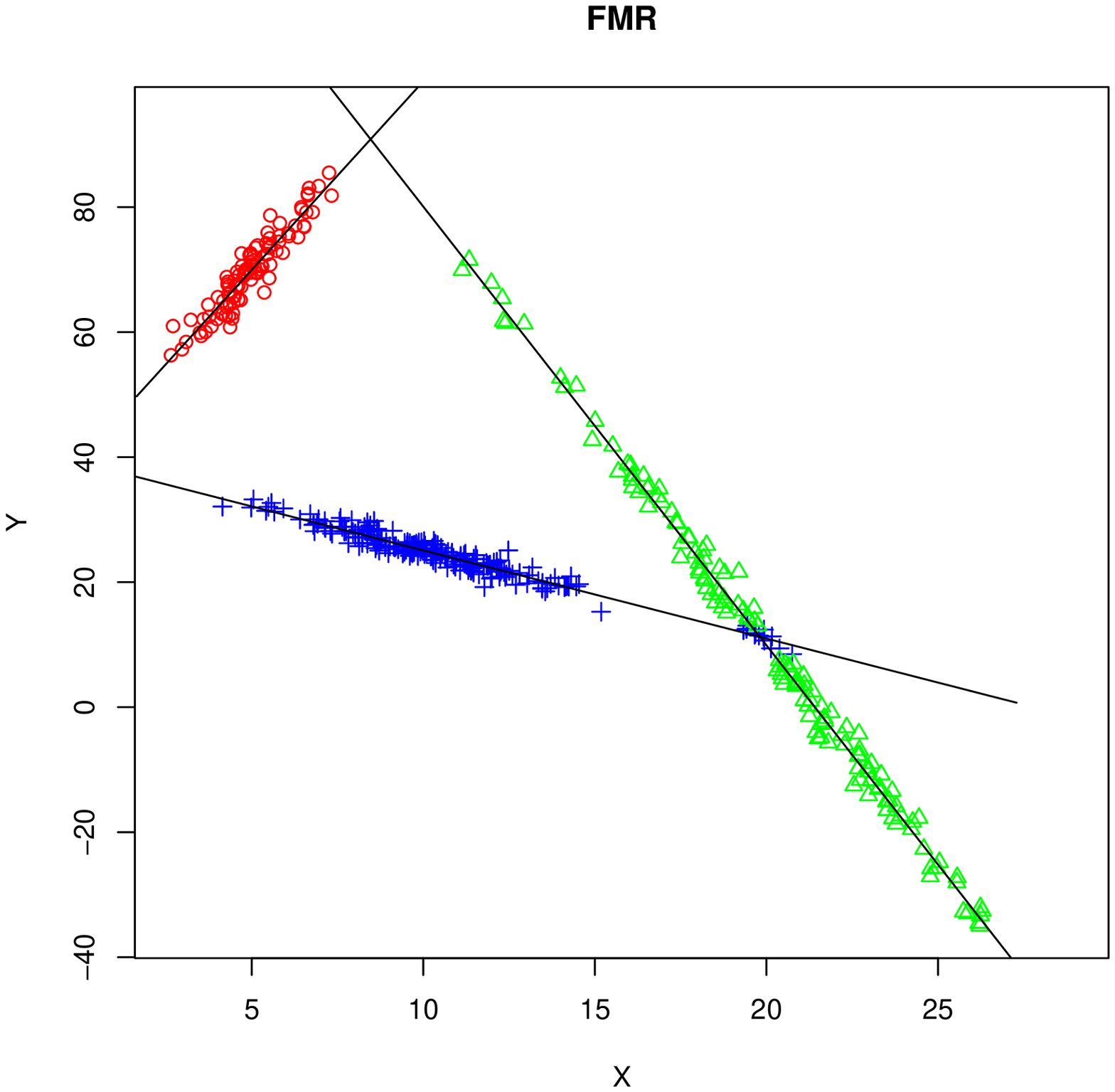}
	\end{center}
	\vspace{-.5 cm}
  \caption{ Example \ref{rettetirng}. Data classification and fitted lines according to CWM and FMR. The symbol $+$ denotes data classified in group 1,  $\circ$ denotes 
  data classified in group 2 and $\triangle$ denotes data classified in group 3.}\label{fig:rettetriang}
\end{figure}
}\end{Ex}

\begin{Ex}\label{allin}{\rm
In the third example, we consider $G=3$ groups with the same  conditional distributions; the data have been generated based on the following parameters:
\begin{center}
\begin{tabular}{ccrcccrc}
\hline\hline
				& & \multicolumn{2}{c}{$\phi(x;\mu_i, \sigma^2_g)$} & \qquad & \multicolumn{3}{c}{$\phi(y; b_{g0}+b_{g1}x,   \sigma^2_{\varepsilon, g})$}  \\ \cline{3-4} \cline{6-8}
				& $N_g$ & $\mu_g$ & $\sigma_g$ & & $b_{g0}$ & $b_{g1}$ & $\sigma_{\epsilon,g}$ \\ \hline
{\em Group 1} & 100 &   5  & 2 & & 2 & 6 & 2 \\ 
{\em Group 2} & 200 & $20$ & 1 & & 2 & 6 & 1 \\
{\em Group 3} & 150 & $40$ & 2& & 2 & 6 & 2 \\
\hline\hline
\end{tabular}
\end{center}
Data are shown in Figure \ref{fig:allin}.
 The obtained results according to  CWM, FMR and FMRC are summarized in the following table:
\begin{center}
\begin{tabular}{cccc}
\hline\hline
  & $\Lambda$ & $\cE$ & $\eta$ \\ \hline
CWM &   0.0146 & 1.678 &  0.00\%\\
FMR &   0.9782 & 2.25 & 51.33\% \\
FMRC & 0.9581 & 0.87 & 45.78\% \\
\hline\hline
\end{tabular}
\end{center}
As we can see, CWM perfectly identifies the three groups, while  both FMR and FMRC lead to very bad results because the groups have the same conditional distributions. As a matter of fact, the reason is that the classification of CWM is based on both marginal and conditional distributions. The scatter plots of data classified according to CWM, FMR and FMRC  are given 
in Figure \ref{fig:allin}. 
\begin{figure}[h] 
	\begin{center}
		\includegraphics[height=6cm, width=6 cm]{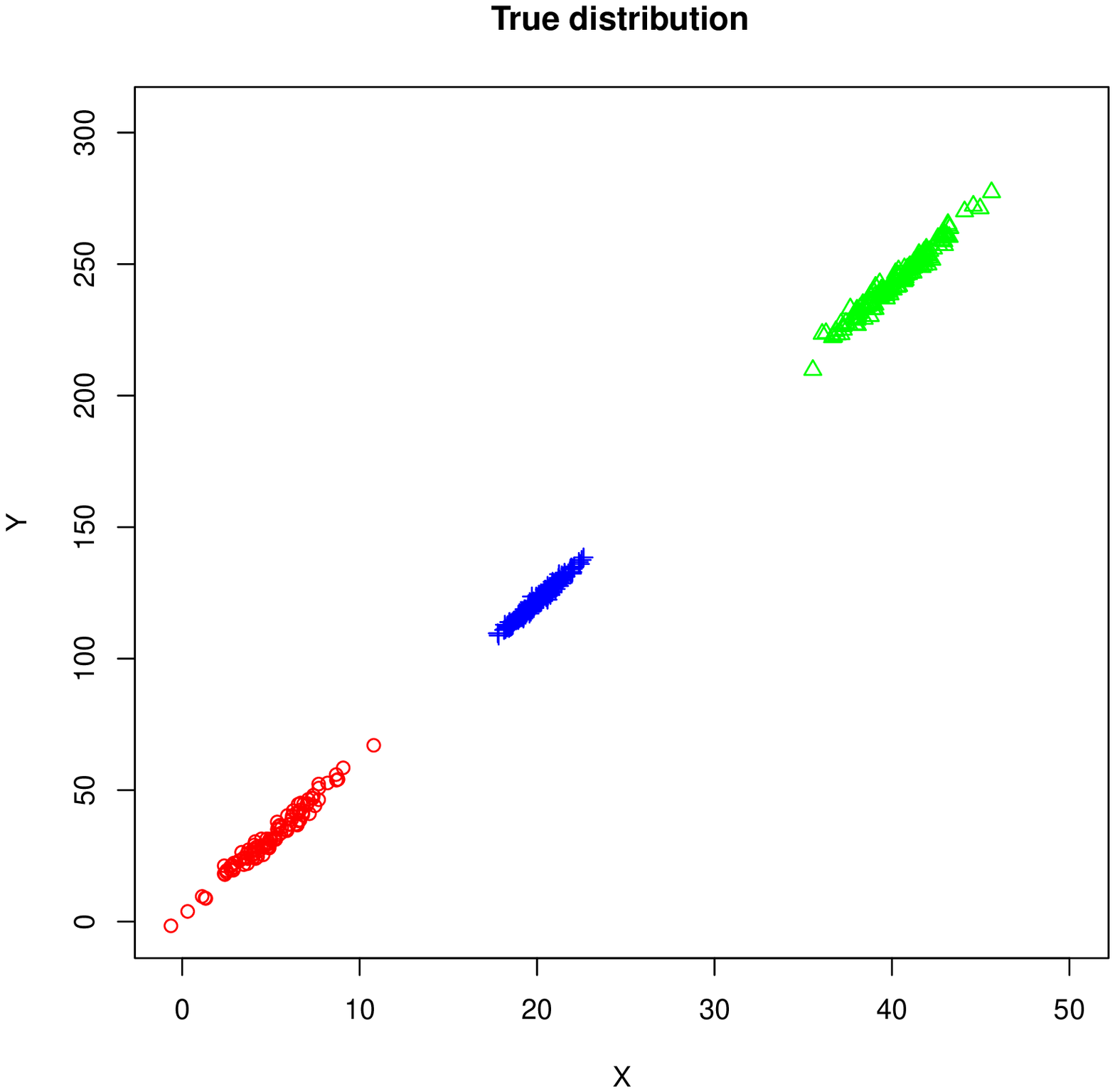}
		\includegraphics[height=6cm, width=6 cm]{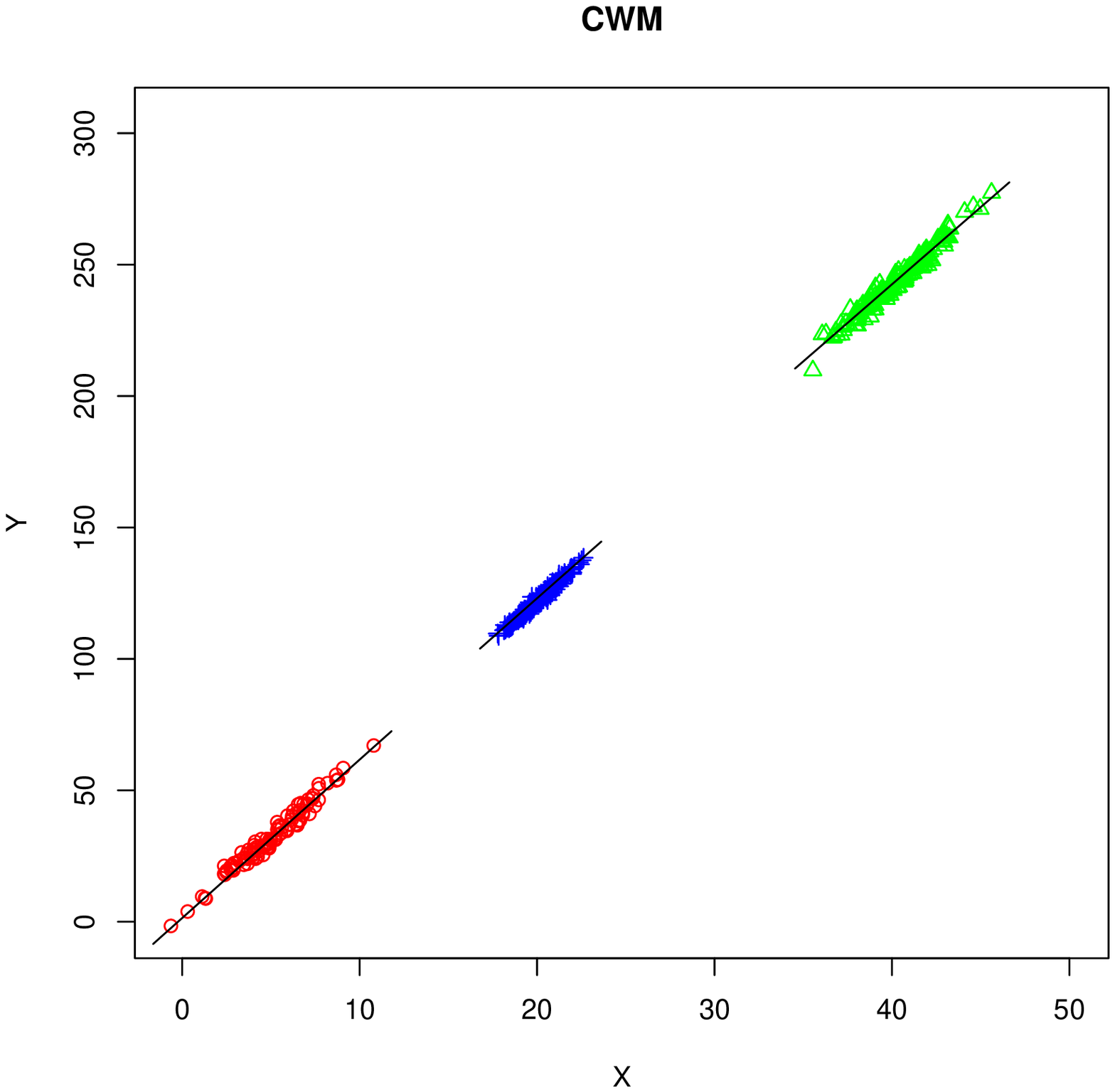} \\
		\includegraphics[height=6cm, width=6 cm]{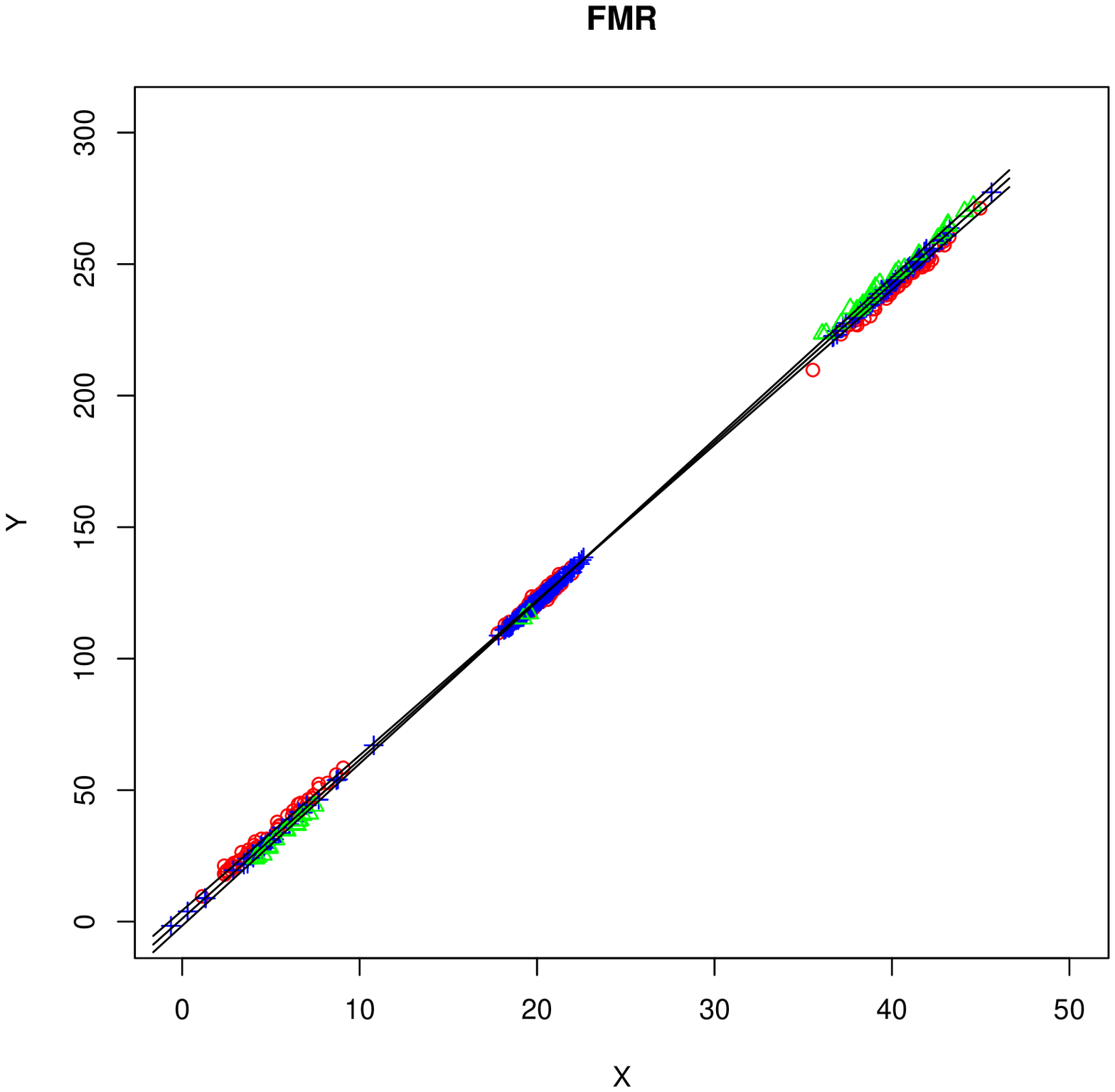}
		\includegraphics[height=6cm, width=6 cm]{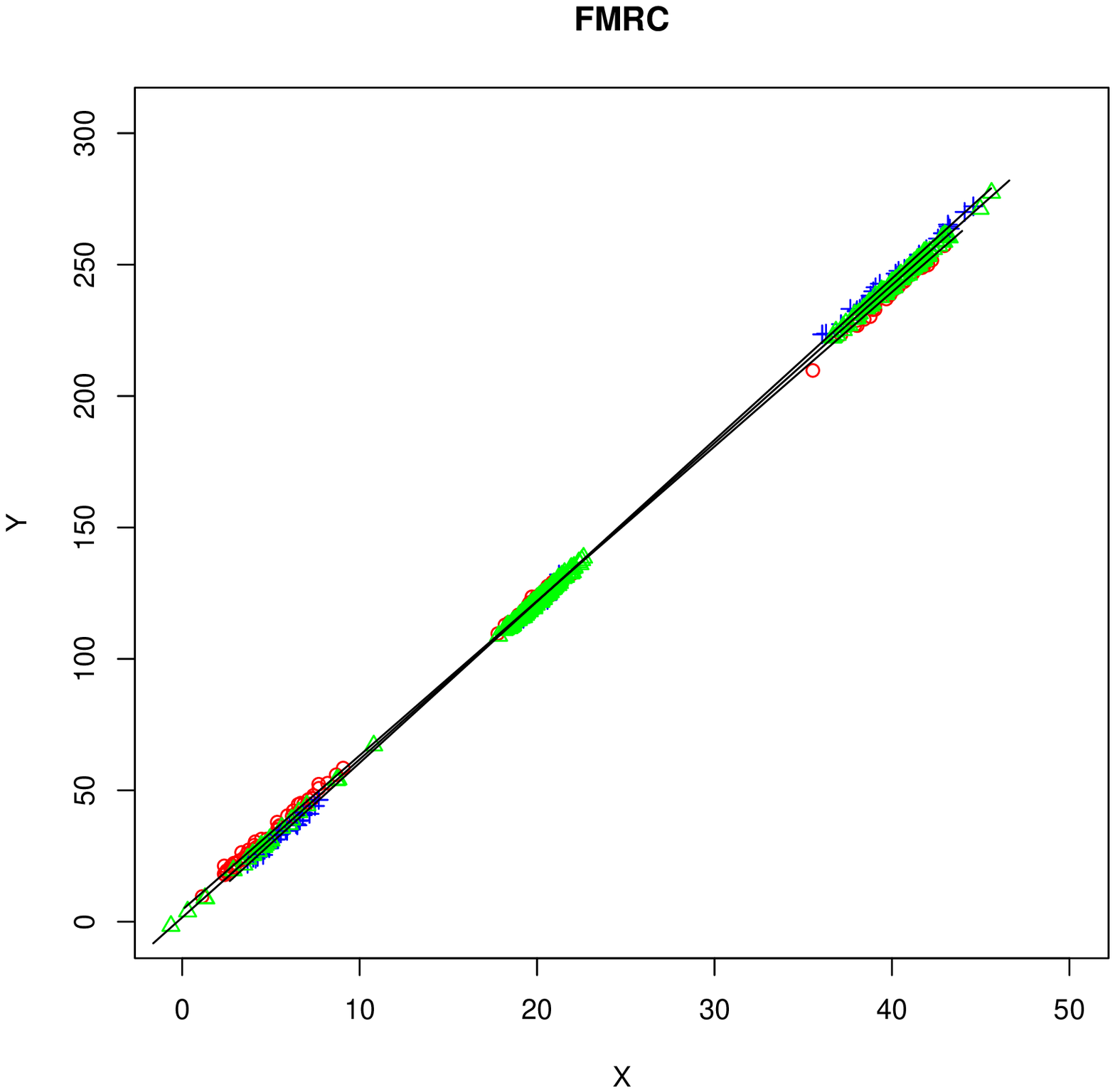}
	\end{center}
	\vspace{-.5 cm}
  \caption{ Example \ref{allin}. True distribution and data classification and fitted lines according to CWM, FMR and FMRC. The symbol $+$ denotes data classified in group 1,  $\circ$ denotes 
  data classified in group 2 and $\triangle$ denotes data classified in group 3.}\label{fig:allin}
\end{figure}
}\end{Ex}

\subsection{Robust clustering of noisy  data}\label{sec:noisydata}
In this section we present the results of some numerical analysis concerning robust clustering from noisy simulated data.
The data have been fitted according to a procedure based on three steps. First, we identify a subset $\cO$ of units which are marked as outliers (i.e. noise data); secondly, we model the reduced dataset $\cD'=\cD \setminus \cO$ using CWM and estimate the parameters. Finally, based on such estimate, we classify the whole dataset $\cD$ into $G$ groups plus a group of noise data.

The first step can be performed following different strategies.  Once the estimates of the parameters of the $g$-th group ($g=1,\ldots, G$), have been obtained,  consider the squared Mahalanobis distance between each unit and the $g$-th local estimate.  
In the framework of robust clustering via FMT, \cite{PeMcLa:00} proposed an approach based on the maximum likelihood; in particular, an observation $\bx_n$ is treated as an outlier (and thus it will be classified as noise data) if 
\begin{equation*}
\sum_{g=1}^G \hat{z}_{jn} \delta(\bx_n; \hat{\bmu}_g, \hat{\bSigma}_g) > \chi^2_{1-\alpha}(q) \, ,  \label{outdec}
\end{equation*}
where $\hat{z}_{jn}=1$ if  $\bx_n$ unit is classified in the $j$-th group according to the maximum posterior probability 
and 0 otherwise, $\delta(\bx_n; \hat{\bmu}_g, \hat{\bSigma}_g) = (\bx_n - \hat{\bmu}_g)'\hat{\bSigma}_g^{-1}(\bx_n - \hat{\bmu}_g)$ and 
  $\chi^2_{1-\alpha}(q)$ denotes the quantile of order $(1-\alpha)$ of the chi-squared distribution 
with $q$ degrees of freedom. More recent approaches are based on the forward search, see e.g.  \cite{RCAPT:08}, \cite{RAC:09}, maximum likelihood estimation with a trimmed sample, see \citet{CMM:08}, \citet{GaRi:09}, \citet{GaRi:09b}
and on multivariate outlier tests based on the minimum covariance determinant estimator, see \cite{Cer:10}.

For the scope of the present paper, we followed \cite{PeMcLa:00}'s strategy, using a Student-$t$ CWM. Cluster Weighted Modeling of noisy data according to  other strategies provides ideas for further research.

As far as the second step is concerned, the parameters have been estimated once the reduced dataset $\cD'=\cD \setminus \cO$ according to either Gaussian or \textit{Student}-$t$ CWM
has been obtained.
In the following, such strategies will be referred to as {\em Student-Gaussian CWM} ({\em t}G-CWM) and {\em Student-Student CWM} ({\em tt}-CWM), respectively.
Finally the data have been classified into $G+1$ groups. A similar strategy has also been considered in \cite{GrIn:10}.
\begin{Ex}[{\em Gaussian simulated data with noise.}]\label{ex:LG1}{\rm
 The first simulated dataset  concerns a sample
of 300 units generated according to  (\ref{CWM-GGlin}) with $G=3$, $d=1$,  $\pi_1=\pi_2=\pi_3=1/3$.
The parameters are listed in the following table for two different values of  $\sigma=\sigma_{\epsilon}=2$ and $\sigma=\sigma_{\epsilon}=4$
\begin{center}
\begin{tabular}{ccrcccrc}
\hline\hline
				& & \multicolumn{2}{c}{$\phi(x;\mu_i, \sigma^2_g)$} & \qquad & \multicolumn{3}{c}{$\phi(y; b_{g0}+b_{g1}x,   \sigma^2_{\varepsilon, g})$}  \\ \cline{3-4} \cline{6-8}
				& $N_g$ & $\mu_g$ & $\sigma_g$ & & $b_{g0}$ & $b_{g1}$ & $\sigma_{\epsilon,g}$ \\ \hline
{\em Group 1} & 100 &   5  & $\sigma$ & & 40 & $6$ & $\sigma_{\epsilon}$ \\ 
{\em Group 2} & 100 & $10$ & $\sigma$ & & 40 & $-1.5$ & $\sigma_{\epsilon}$ \\
{\em Group 3} & 100 & $20$ & $\sigma$ & & 150 & $-7$ & $\sigma_{\epsilon}$ \\
\hline\hline
\end{tabular}
\end{center}
The sample data $\{(x_n,y_n) \}_{n=1, \ldots, 300}$ have been obtained as follows: first, we have generated the samples 
$\bx_1,\ldots,\bx_N$   according to $G=3$ Gaussian distributions with parameters $(\mu_g,\sigma_g)$, $g=1,...,G$. Afterwards, for each $x_g$ we  generated the value $y_g$ (corresponding to the $Y$-variable)  according to a Gaussian distribution with mean $b_{g0}+b_{g1}x$ and variance $\sigma^2_{\epsilon}$. 
Then, the above data set has been augmented by including a sample of 50 points generated with a uniform distribution in the rectangle $[-5,30]\times[-50,130]$ in order to simulate noise. Thus, the whole dataset $\cD$  contains $N=350$ units, 
see  Figure \ref{fig:LG1n}. 
\begin{figure}[h] 
	\begin{center}
	\begin{tabular}{cc}
		\includegraphics[height=6 cm, width=6 cm]{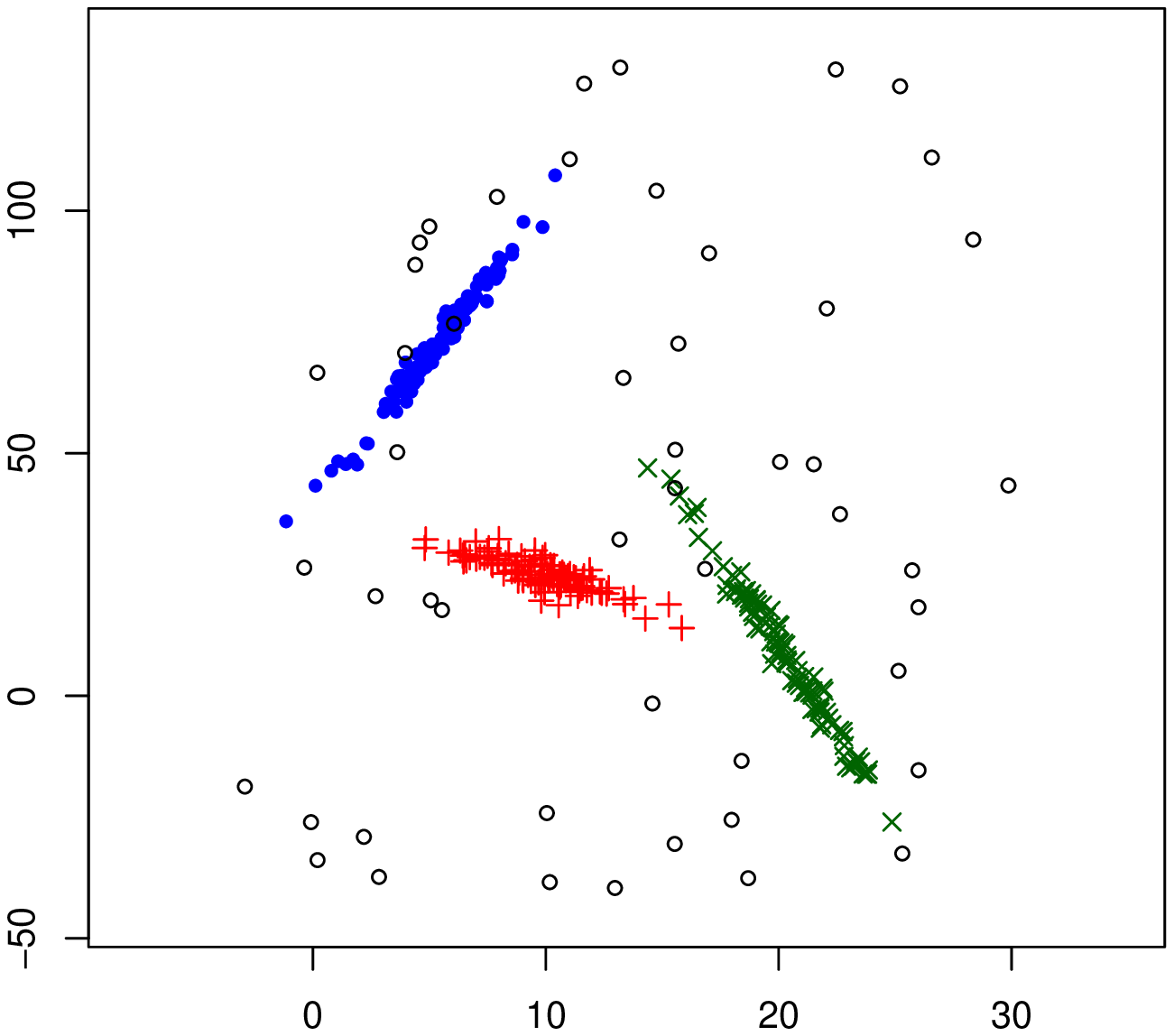} &
		\includegraphics[height=6 cm, width=6 cm]{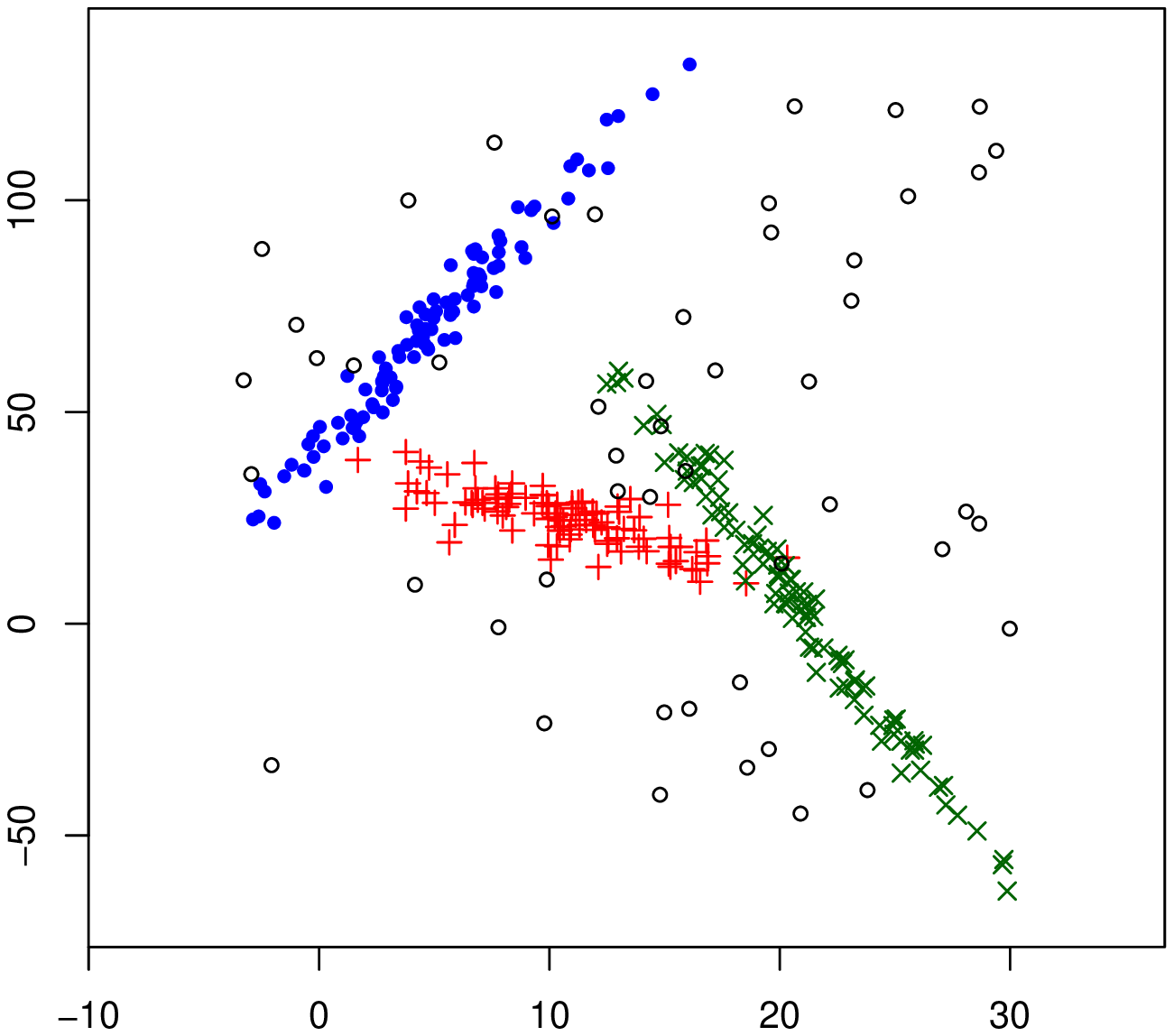} \vspace{-5mm} \\ 
a) & b) \\
\end{tabular}
	\end{center}
  \caption{\textit {\rm Example \ref{ex:LG1}: a) data with $\sigma=2$, b) data with $\sigma=4$ (circles represent noise)}}\label{fig:LG1n}
\end{figure}

The results have been summarized in Table \ref{tab:ex1n}. {\em t}G-CWM and {\em tt}-CWM have in practice the same performance. In the case $\sigma=2$, {\em tt}-CWM slightly outperformed {\em t}G-CWM (the misclassification rates were $\eta=6.00\%$ and $\eta=5.71\%$, respectively); however, {\em tt}-CWM recognized a larger number of outliers than {\em t}G-CWM, viceversa in the case $\sigma=4$  we observed $\eta=4.29\%$ and $\eta=5.71\%$, respectively. We remark that the smallest misclassification error $\eta$ corresponds to the model with the smallest mean squared error $\cE$. 
\begin{center}
\begin{table}[h]
\begin{tabular}{ccc}
a) Student-Gaussian CWM & & \\
\begin{tabular}{c|cccc}
\hline\hline
& \multicolumn{3}{c}{estimated} \\ \cline{2-5}
true & 1 & 2 & 3 & outlier \\ \hline
1 & 98 & 0 & 0 & 2 \\
2 & 0  & 97 & 0 & 3 \\
3 & 0 & 0 & 100 & 0 \\
outlier & 1 & 0 & 15 & 34 \\
\hline\hline
\end{tabular}
& \hspace{1cm} &
\begin{tabular}{c|cccc}
\hline\hline
& \multicolumn{3}{c}{estimated} \\ \cline{2-5}
true & 1 & 2 & 3 & outlier \\ \hline
1 & 99 & 0 & 0 & 1 \\
2 & 0  & 98 & 1 & 1 \\
3 & 0 & 1 & 99 & 0 \\
outlier & 5 & 2 & 4 & 39 \\
\hline\hline
\end{tabular} \\ \\
case $\sigma=2$: $\cE=7.97, \eta=6.00\%$ & & case $\sigma=4$: $\cE=4.34, \eta=4.29\%$ \\ \\
b) Student-Student CWM & & \\
\begin{tabular}{c|cccc}
\hline\hline
& \multicolumn{3}{c}{estimated} \\ \cline{2-5}
true & 1 & 2 & 3 & outlier \\ \hline
1 & 94 & 0 & 0 & 6 \\
2 & 0  & 92 & 0 & 8 \\
3 & 0 & 0 & 99 & 1 \\
outlier & 1 & 0 & 4 & 45 \\
\hline\hline
\end{tabular}
& \hspace{1cm} &
\begin{tabular}{c|cccc}
\hline\hline
& \multicolumn{3}{c}{estimated} \\ \cline{2-5}
true & 1 & 2 & 3 & outlier \\ \hline
1 & 98 & 0 & 0 & 2 \\
2 & 0  & 93 & 4 & 3 \\
3 & 0 & 0 & 100 & 0 \\
outlier & 4 & 0 & 7 & 39 \\
\hline\hline
\end{tabular} \\ \\
case $\sigma=2$: $\cE=2.97, \eta=5.71\%$ & & case $\sigma=4$: $\cE=5.8, \eta=5.71\%$ \\ \\
\end{tabular}
\caption{Summary of the results concerning Example \ref{ex:LG1}: confusion matrices, mean squared error and misclassification rate for data fitting using both
Student-Gaussian CWM and Student-Student CWM. The smallest misclassification error has been attained in correspondence with the smallest value of $\cE$.}\label{tab:ex1n}
\end{table}
\end{center}
}\end{Ex}

\begin{Ex}[{\em Gaussian simulated data with noise.}]\label{ex:LG2}{\rm
The second simulated example concerns a data set  of size $150$ generated according to  model (\ref{CWM-GGlin}) with $G=3$, $d=1$,  $\pi_1=\pi_2=\pi_3=1/3$. 
The parameters are listed in the following table for two different values of  $\sigma=\sigma_{\epsilon}=2$ and $\sigma=\sigma_{\epsilon}=4$:
\begin{center}
\begin{tabular}{ccrcccrc}
\hline\hline
				& & \multicolumn{2}{c}{$\phi(x;\mu_i, \sigma^2_g)$} & \qquad & \multicolumn{3}{c}{$\phi(y; b_{g0}+b_{g1}x,   \sigma^2_{\varepsilon, g})$}  \\ \cline{3-4} \cline{6-8}
				& $N_g$ & $\mu_g$ & $\sigma_g$ & & $b_{g0}$ & $b_{g1}$ & $\sigma_{\epsilon,g}$ \\ \hline
{\em Group 1} & 100 &   5  & $\sigma$ & & 2 & $6$ & $\sigma_{\epsilon}$ \\ 
{\em Group 2} & 100 & $10$ & $\sigma$ & & 2 & $-1.5$ & $\sigma_{\epsilon}$ \\
{\em Group 3} & 100 & $40$ & $\sigma$ & & 2 & $-7$ & $\sigma_{\epsilon}$ \\
\hline\hline
\end{tabular}
\end{center}
i.e. the data are divided into $G=3$ groups along one straigth line.
Afterwards, we added to the previous data a sample of 25 points generated by a uniform distribution in the rectangle $[-5,30]\times[-50,130]$ in order to simulate noise. 
Thus,  $\cD$  contains $N=175$ units,  see  Figure \ref{fig:LG2n}. 

The results have been summarized in Table \ref{tab:ex2n}. In the case $\sigma=2$, {\em t}G-CWM slightly outperformed {\em tt}-CWM, the misclassification rates were $\eta=4.00\%$ and $\eta=5.14\%$, respectively; in the case $\sigma=4$ {\em t}G-CWM essentially identifies two groups (and thus $\eta=40\%$), while {\em tt}-CWM recognized the three groups with a misclassification rate $\eta=8.00\%$. 
Figure \ref{fig:LG2n}b) explains the reason for the relevant misclassification error in data fitting via {\em t}G-CWM: as a matter of fact two clusters are very close; in this case, {\em t}G-CWM identifies such two clusters as a whole, while {\em tt}-CWM correctly separates them.
We point out that also in this case the smallest misclassification error $\eta$ corresponds to the model with the smallest mean squared error $\cE$. 

\begin{center}
\begin{table}[h]
\begin{tabular}{ccc}
a) Student-Gaussian CWM & & \\
\begin{tabular}{c|cccc}
\hline\hline
& \multicolumn{3}{c}{estimated} \\ \cline{2-5}
true & 1 & 2 & 3 & outlier \\ \hline
1 & 47 & 0 & 0 & 3 \\
2 & 0  & 50 & 0 & 1 \\
3 & 0 & 0 & 49 & 1 \\
outlier & 0 & 2 & 1 & 22 \\
\hline\hline
\end{tabular}
& \hspace{1cm} &
\begin{tabular}{c|cccc}
\hline\hline
& \multicolumn{3}{c}{estimated} \\ \cline{2-5}
true & 1 & 2 & 3 & outlier \\ \hline
1 & 0 & 50 & 0 & 0 \\
2 & 4  & 50 & 0 & 0 \\
3 & 0 & 0 & 50 & 4 \\
outlier & 19 & 0 & 1 & 5 \\
\hline\hline
\end{tabular} \\ \\
case $\sigma=2$: $\cE=2.29, \eta=4.00\%$ & & case $\sigma=4$: $\cE=71.39, \eta=40.00\%$ \\ \\
b) Student-Student CWM & & \\
\begin{tabular}{c|cccc}
\hline\hline
\multicolumn{4}{c}{estimated} \\ \cline{2-5}
true & 1 & 2 & 3 & outlier \\ \hline
1 & 46 & 0 & 0 & 4 \\
2 & 0  & 49 & 0 & 1 \\
3 & 0 & 0 & 48 & 2 \\
outlier & 0 & 1 & 1 & 23 \\
\hline\hline
\end{tabular}
& \hspace{1cm} &
\begin{tabular}{c|cccc}
\hline\hline
& \multicolumn{3}{c}{estimated} \\ \cline{2-5}
true & 1 & 2 & 3 & outlier \\ \hline
1 & 49 & 0 & 0 & 1 \\
2 & 4  & 46 & 0 & 0 \\
3 & 0 & 0 & 46 & 4 \\
outlier & 0 & 0 & 5 & 20 \\
\hline\hline
\end{tabular} \\ \\
case $\sigma=2$: $\cE=7.25, \eta=5.14\%$ & & case $\sigma=4$: $\cE=31.96, \eta=8.00\%$ \\ \\
\end{tabular}
\caption{Summary of the results concerning Example \ref{ex:LG2}: confusion matrices, mean squared error and misclassification rate for data fitting using both
Student-Gaussian CWM and Student-Student CWM. The smallest misclassification error is obtained corresponding to the smallest value of $\cE$.}\label{tab:ex2n}
\end{table}
\end{center}

\begin{figure}[h] 
	\begin{center}
	\begin{tabular}{cc}
		\includegraphics[height=6 cm, width=6 cm]{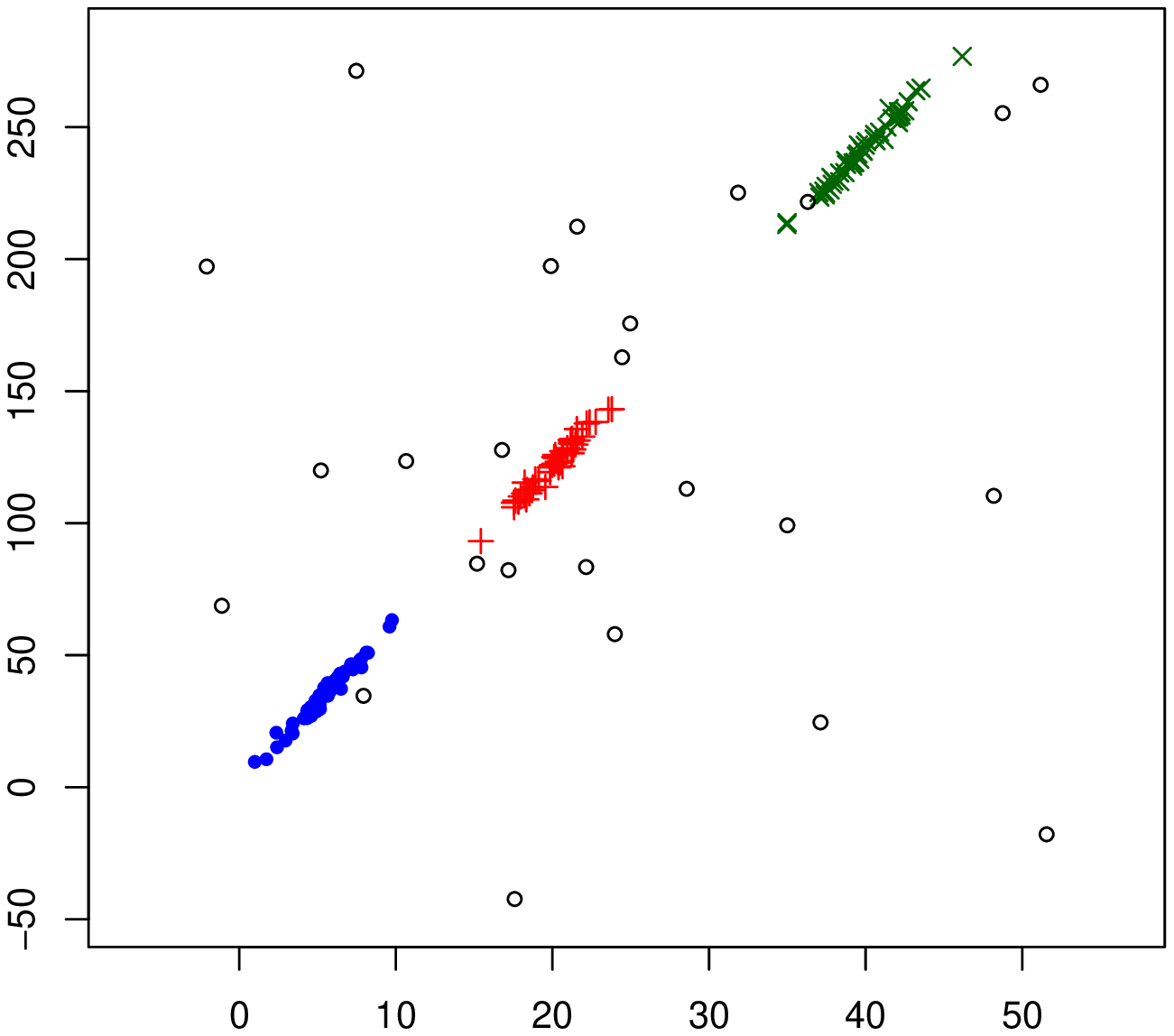} &
		\includegraphics[height=6 cm, width=6 cm]{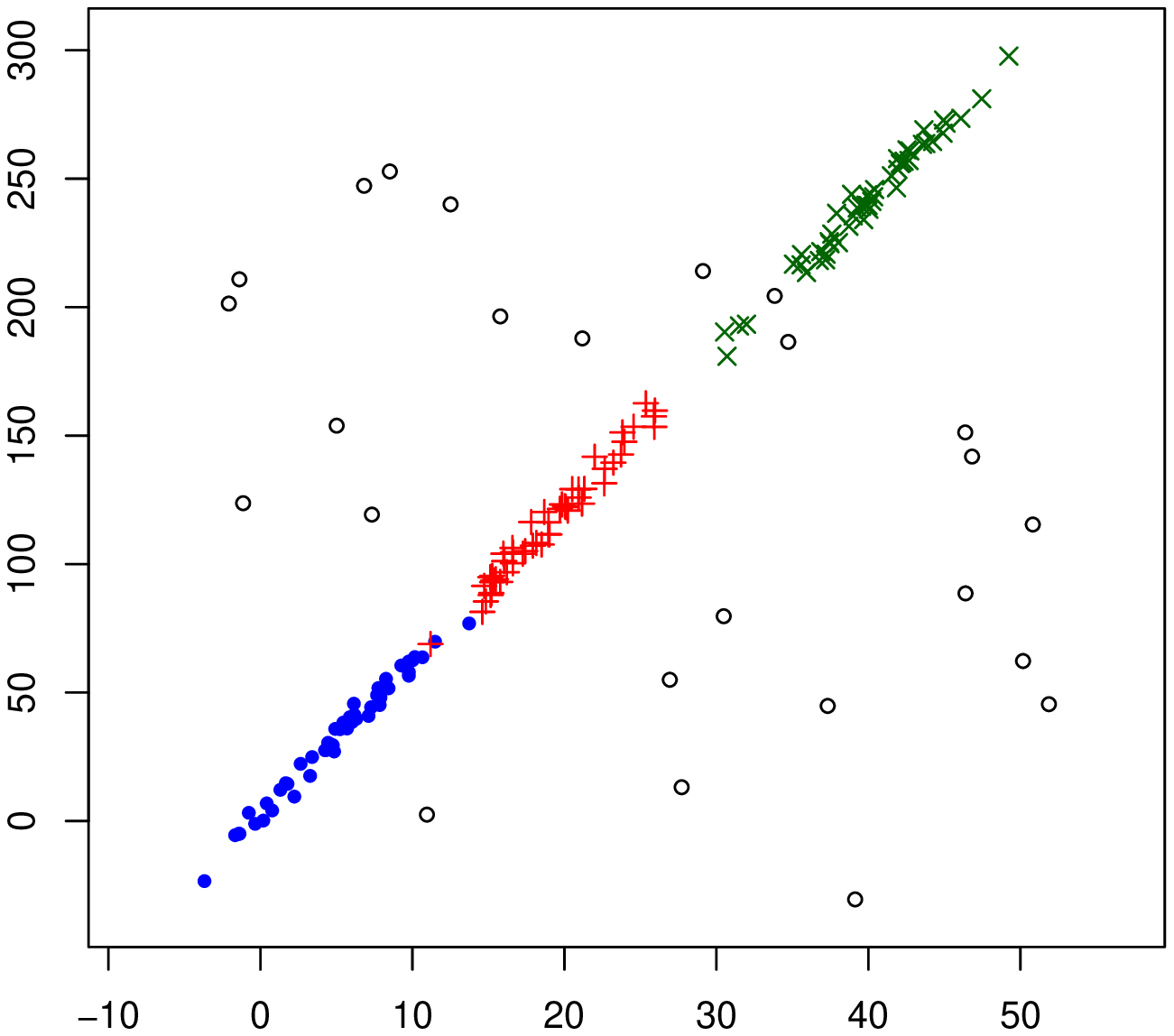} \vspace{-5mm} \\ 
a) & b) \\
\end{tabular}
	\end{center}
  \caption{\textit {\rm Example \ref{ex:LG2}: a) data with $\sigma=2$, b) data with $\sigma=4$ (circles represent noise)}}\label{fig:LG2n}
\end{figure}
}\end{Ex}

\begin{Ex}[{\em Bivariate Linear Gaussian simulated noisy data.}]\label{ex:bivGauss}{\rm
The third example concerns a data set  of size $300$ generated according again to  (\ref{CWM-GGlin}) with $G=2$, $d=2$,  $\pi_1=\pi_2=1/2$ with  the following parameters for $p(y|\bx,\Omega_g)$:
\[ \mu(\bx, \bbeta_1) = 6 x_1+ 1.2 x_2 \qquad {\rm and} \qquad \mu(\bx, \bbeta_2)= -1.5 x_1 + 3 x_2 \]
that is $\bmu_1= (6,1.2)'$ and $\bmu_2 = (-1.5, 3)'$,  and the following parameters for $p(\bx|\Omega_g)=\phi_2(\bx; \bmu_g, \bSigma_g)$, $g=1,2$, for two different values of $\sigma_1= \sigma_2  =\sigma$ and $\sigma_{\epsilon,1}  = \sigma_{\epsilon,2} = \sigma_{\epsilon}$, i.e. $\sigma=\sigma_{\epsilon}=2$ and $\sigma=\sigma_{\epsilon}=4$:
\[ \bmu_1= (5,20)' \; , \quad \bSigma_1 = \begin{pmatrix} 4 & -0.1 \\ -0.1 & 4 \end{pmatrix} \qquad {\rm and} \qquad \bmu_2=(2,4) \; , 
\quad \bSigma_2 = \begin{pmatrix} 4 & 0.1 \\ 0.1 & 4 \end{pmatrix} \]

%
Afterwards,  a sample of 50 points generated by a uniform distribution in the rectangle $[-5,40]\times[-5,40]\times[-20,170]$ has been added in order to simulate noise. Thus, the dataset $\cD$  contains $N=350$ units.
The results have been summarized in Table \ref{tab:ex3}. In the case $\sigma=2$, {\em t}G-CWM slightly outperformed {\em tt}-CWM, the misclassification rates were $\eta=2.00\%$ and $\eta=2.29\%$, respectively; similar results we obtained in the case $\sigma=4$, where we got $\eta=6.57\%$ and $\eta=7.43\%$, respectively. 
Again the smallest misclassification error $\eta$ has been attained corresponding to the model with the smallest mean squared error $\cE$.

\begin{center}
\begin{table}[h]
\begin{tabular}{ccc}
a) Student-Gaussian CWM & & \\
\begin{tabular}{c|ccc}
\hline\hline
& \multicolumn{2}{c}{estimated} \\ \cline{2-4}
true & 1 & 2  & outlier \\ \hline
1 & 149 & 0 &  1 \\
2 & 0  & 144 &  6 \\
outlier & 0 & 0  & 50 \\
\hline\hline
\end{tabular}
& \hspace{1cm} &
\begin{tabular}{c|cccc}
\hline\hline
& \multicolumn{2}{c}{estimated} \\ \cline{2-4}
true & 1 & 2  & outlier \\ \hline
1 & 149 & 1 &  0 \\
2 & 0  & 145 &  5 \\
outlier & 0 & 2  & 48 \\
\hline\hline
\end{tabular} \\ \\
case $\sigma=2$: $\cE=2.08, \eta=2.00\%$ & & case $\sigma=4$: $\cE=2.32, \eta=6.57\%$ \\ \\
b) Student-Student CWM & & \\
\begin{tabular}{c|ccc}
\hline\hline
& \multicolumn{2}{c}{estimated} \\ \cline{2-4}
true & 1 & 2  & outlier \\ \hline
1 & 139 & 0 &  11 \\
2 & 0  & 138 & 12 \\
outlier & 0 & 0  & 50 \\
\hline\hline
\end{tabular}
& \hspace{1cm} &
\begin{tabular}{c|cccc}
\hline\hline
& \multicolumn{2}{c}{estimated} \\ \cline{2-4}
true & 1 & 2  & outlier \\ \hline
1 & 140 & 1 &  9 \\
2 & 0  & 135 &  15 \\
outlier & 0 & 1  & 49 \\
\hline\hline
\end{tabular} \\ \\
case $\sigma=2$: $\cE=3.94, \eta=2.29\%$ & & case $\sigma=4$: $\cE=4.64, \eta=7.43\%$ \\ \\
\end{tabular}
\caption{Summary of the results concerning Example \ref{ex:bivGauss} (data with noise): confusion matrices, mean squared error and misclassification rate for data fitting using both
Student-Gaussian CWM and Student-Student CWM. The smallest misclassification error is obtained corresponding to the smallest value of $\cE$.}\label{tab:ex3}
\end{table}
\end{center}
}\end{Ex}
\subsection{A comparison between  linear $t$-CWM and FMT in robust clustering}\label{sec:t-compare}
\begin{figure}[h] 
	\begin{center}\hspace{-1 cm}
		\includegraphics[height=8 cm, width=8 cm]{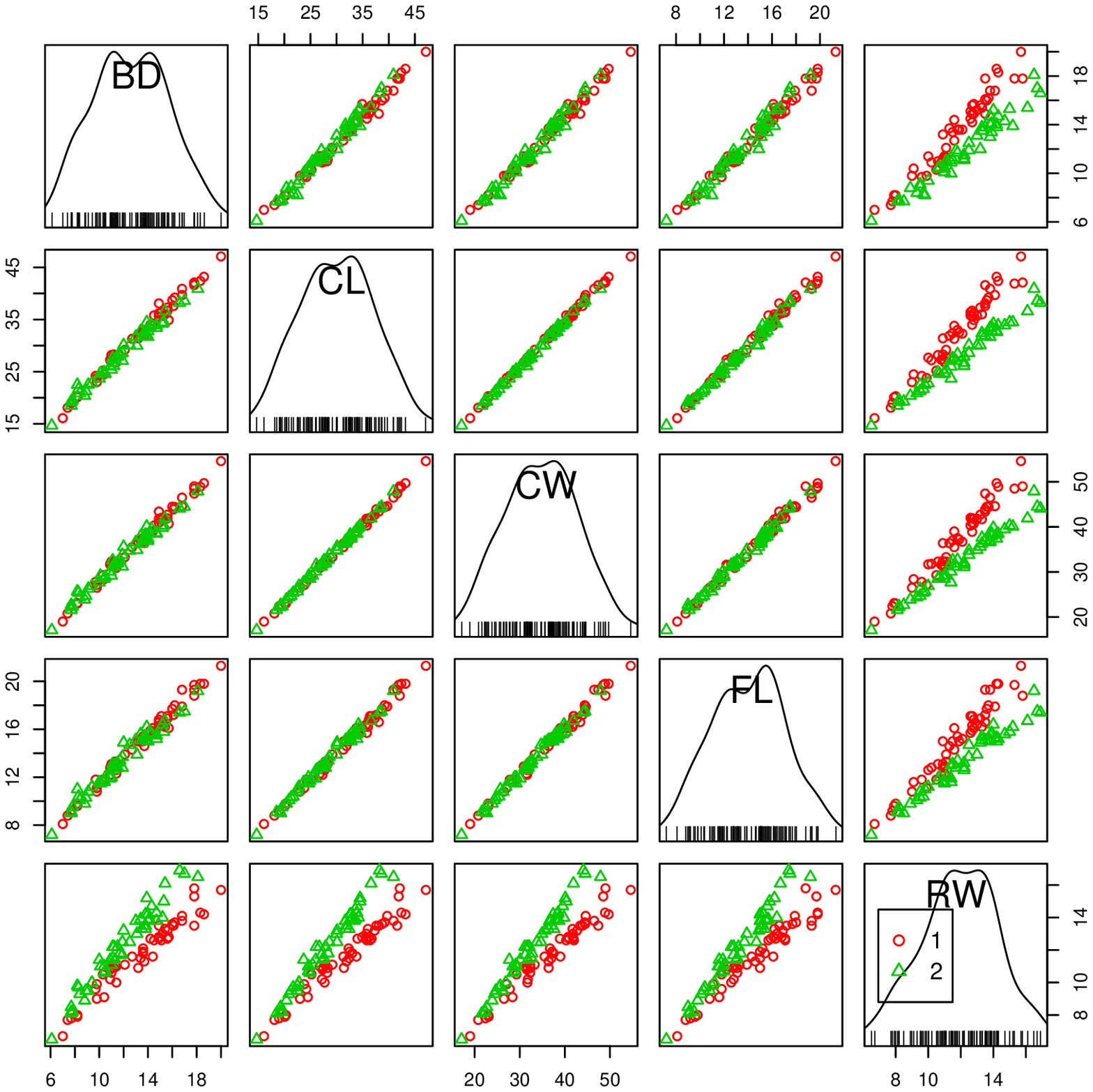} 
	\end{center}
	\vspace{-0.5cm}
  \caption{\em Scatterplot matrix of the crab data set. }\label{fig:crabsdata}
\end{figure}

In this section we present the results concerning robust classification via FMT and \textit{linear $t$}-CWM (\ref{CWM-ttlin}) based on a real data set studied in \citet{CaMa74} about rock crabs of the genus {\em Leptograpsus} (available at http://www.stats.ox.ac.uk/pub/PRNN/). Each specimen has five measurements (expressed in $mm$): width of the frontal lip ({\em FL}), rear width ({\em RW}), length along the mid line ({\em CL}) and maximum width ({\em CW}) of the carapace and body depth ({\em BD}); the data are grouped into two classes by sex, see Figure \ref{fig:crabsdata}.
According to the classes of application of CWM introduced in Section \ref{sec:CWMintro}, this case study concerns a direct application of type B; in particular, in  (\ref{CWM-ttlin}) the variable {\em CL} has been selected as the $Y$-variable. 
In the setting of FMT, this data set has been used in \citet{McLaPe:00}, \citet{PeMcLa:00} and in \citet{LLN04}.  According to such references, here we cluster a sample of 100 units (with $n_1=50$ males and $n_2=50$ females) ignoring their true classification. 

The simulations have been performed along the lines of \citet{McLaPe:00}, \citet{PeMcLa:00}.  Outliers have been inserted in the original data set by adding a constant to the second variate of the 25{\em th} data point. 
In Table \ref{tab:crabmod} the overall misclassification error is reported, comparing both FMT and \textit{linear $t$}-CWM error rates obtained in the best case. 
We read Table \ref{tab:crabmod} beginning from the central row, where the initial data set is considered (without any perturbation, as the constant value is null). \textit{Linear} $t$-CWM  outperformed  FMT. Then, scanning the following rows of the table, the value of the constant raises progressively from 5 to 20 $mm$ and the error rate grows  more slowly for FMT (reaching only 20\%), while it remains almost unchanged for CWM. 
\begin{table}[h]
\caption{\em Comparison of error rates when fitting FMT and $t$-CWM to the crab data set with outliers. Variable {\em CL} has been selected as the $Y$-variable. }\label{tab:crabmod}
\begin{center}
\begin{tabular}{rcccc}
\hline\hline
 &   FMT  &   \textit{linear} $t$-CWM  \\ 
 Constant &  Error Rate &  Error Rate   \\ \hline 
$-15$  & 19\%    & 13\% \\ 
$-10$  & 19\% &  13\% \\
$-5$   & 20\% &  13\% \\
0 &  18\% &   12\% \\
5 &  20\% &   12\% \\
10  & 20\%  & 11\% \\
15 & 20\%  & 12\%\\
20 & 20\%   & 12\% \\
\hline\hline
\end{tabular}
\end{center}
\end{table}
Finally, we remark that misclassification error rates obtained via \textit{linear $t$}-CWM are equivalent to those obtained in \citet{GrIn:10} using suitable constraints on the eigenvalues of the covariance matrices of the two groups.

\subsection{A case study in the medical area}\label{sec:realdata}
\begin{Ex}[Plasma Concentration of Beta-Carotene]
{\rm  
The last example concerns a case study based on real data about Beta-Carotene plasma levels described in \cite{Nier:89}. The data have been recently modeled in  \cite{Schla:09} by using FMR and here we compare such approach with CWM. 
Our analysis concentrates on the relationship between Beta-Carotene plasma level and the amount of Beta-Carotene in the diet using a subset of $N=144$ individuals, with status `male' and `never smoker'. Moreover, following the schema in \cite{Schla:09}, we considered $G=4$ groups. 

To begin with, we remark that the BIC criterion assumes a slightly smaller value for CWM (4298.6) rather than FMR (4324.5), see also Table \ref{tab:beta} (we can compare the two values because FMR can be regarded as a nested model of CWM, see Section \ref{sec:CWM-FMR}). 
Moreover, Gaussian CWM leads to better  separated groups ($\Lambda=0.0934$) than FMR ($\Lambda=0.2655$); on the contrary, the index of weighted model fitting for CWM ($\cE=70.3925$) attains a slightly worse value than FMR ($\cE=54.1186$); however, considering the range of values of Beta-Carotene plasma level, this difference is not relevant and the fitting may be considered  good in both cases.
Figure \ref{fig:dataplasma_G4} shows the classifications obtained by means of FMR and Gaussian CWM, respectively.
Finally, the parameter estimates are reported in Table \ref{tab:parest_G4}, where in parenthesis  the standard error of the estimates are given.
FMR classifies individuals into four straight lines along the dietary Beta-Carotene axis (see Figure \ref{fig:dataplasma_G4}). The effect of dietary Beta-Carotene is rather small in the first subpopulation ($b_{11}= 0.0032$), a little greater in the second and fourth subpopulations ($b_{21}=0.0301$ and $b_{41}= 0.0419$), whereas it is much larger in the third subpopulation ($b_{31}=0.2572$). Gaussian CWM produces a different classification, which takes into account the distribution of dietary Beta-Carotene (see Figure \ref{fig:dataplasma_G4}); in particular, a subpopulation with a negative relationship between Beta-Carotene plasma level and dietary Beta-Carotene ($b_{21}=-0.0393$) is identified.  

In order to complete the data analysis, first we remark that  the histogram of data concerning the amount of Beta-Carotene in the diet ($X$-variable)
shows that the population is heterogeneous with respect to the independent variable, see Figure \ref{fig:histXbeta}. This heterogeneity can be captured by CWM (which models the joint distribution) but not by FMR (which models only the conditional distribution). Secondly, we point out that group 2 in CWM exhibits a negative slope ($b_{21}=-0.0393$), while FMR leads to a model with all positive slopes. The identification of a subpopulation which negatively reacts to dietary Beta-Carotene seems to confirm the recent adverse findings about the effect of antioxidants intake on the incidence of lung cancer, see \cite{Schla:09} p.11. This might be a starting point for further investigations in the biomedical area. 
\begin{figure}[ht] 
	\begin{center}
		\includegraphics[height=7 cm, width=7 cm]{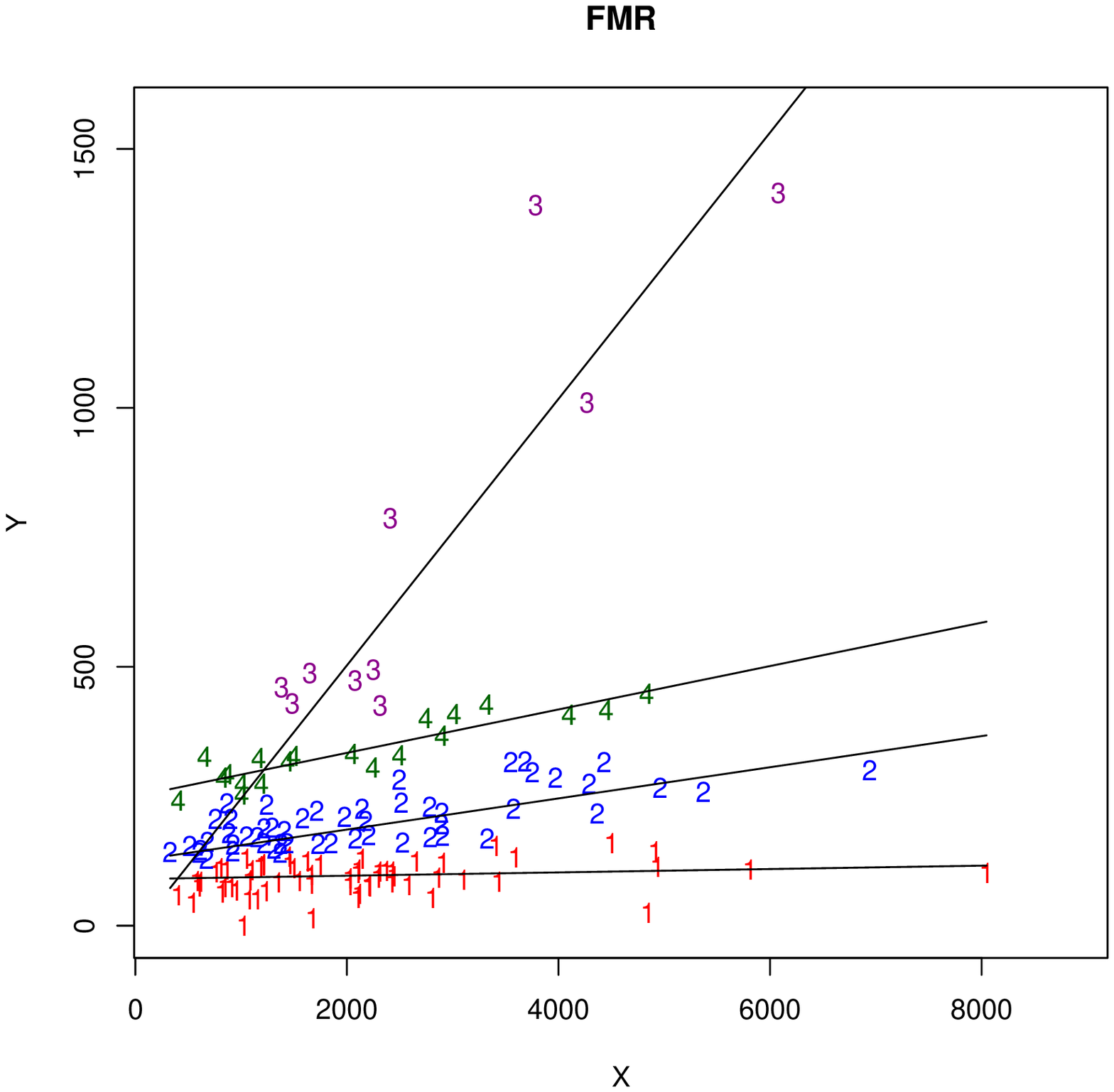}
		\includegraphics[height=7 cm, width=7 cm]{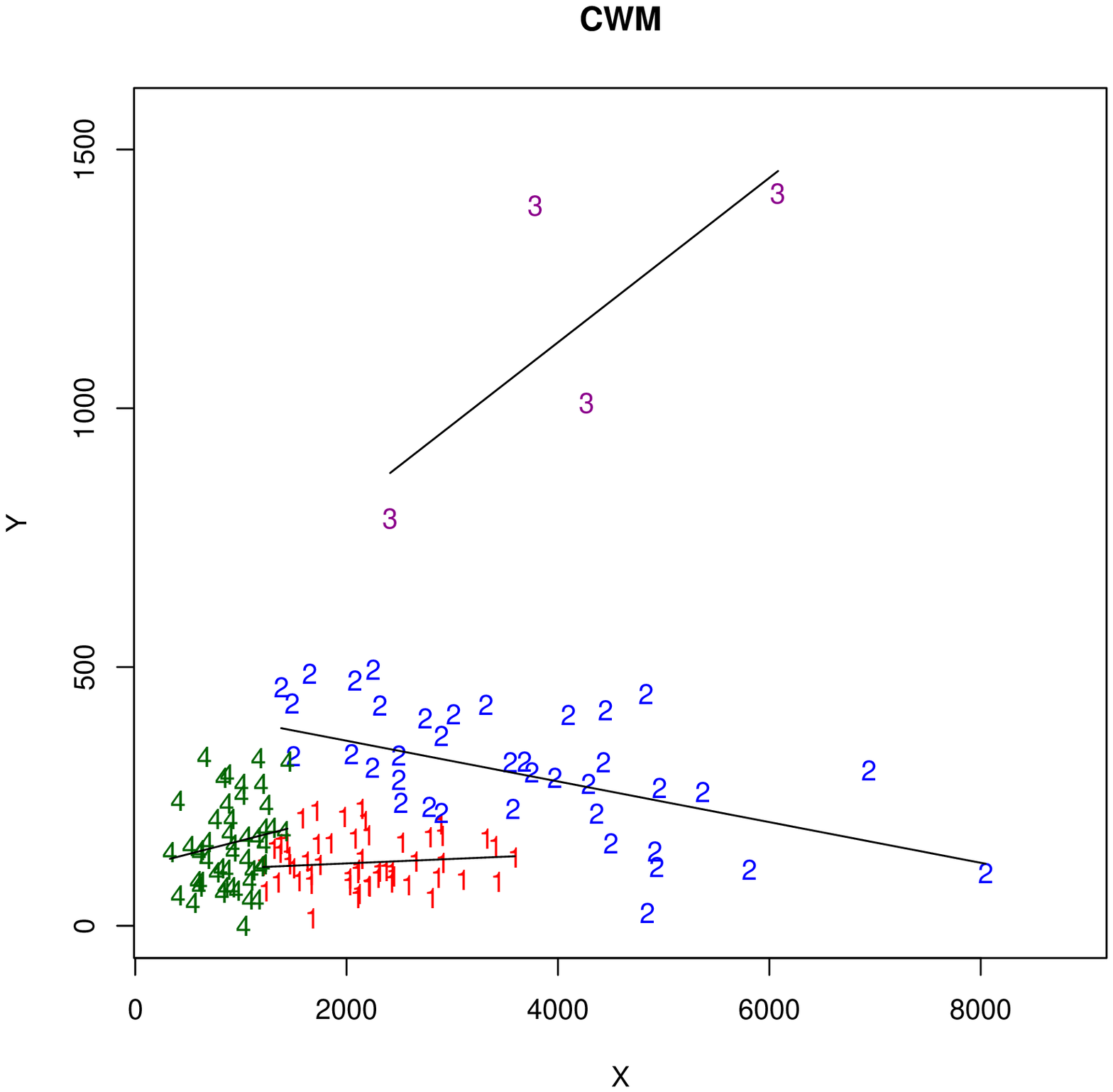}
	\end{center}
  \caption{\rm Betaplasma data: classification according to FMR and CWM.}\label{fig:dataplasma_G4}
\end{figure}

\begin{figure}[ht] 
	\begin{center}
		\includegraphics[height=7 cm, width=7 cm]{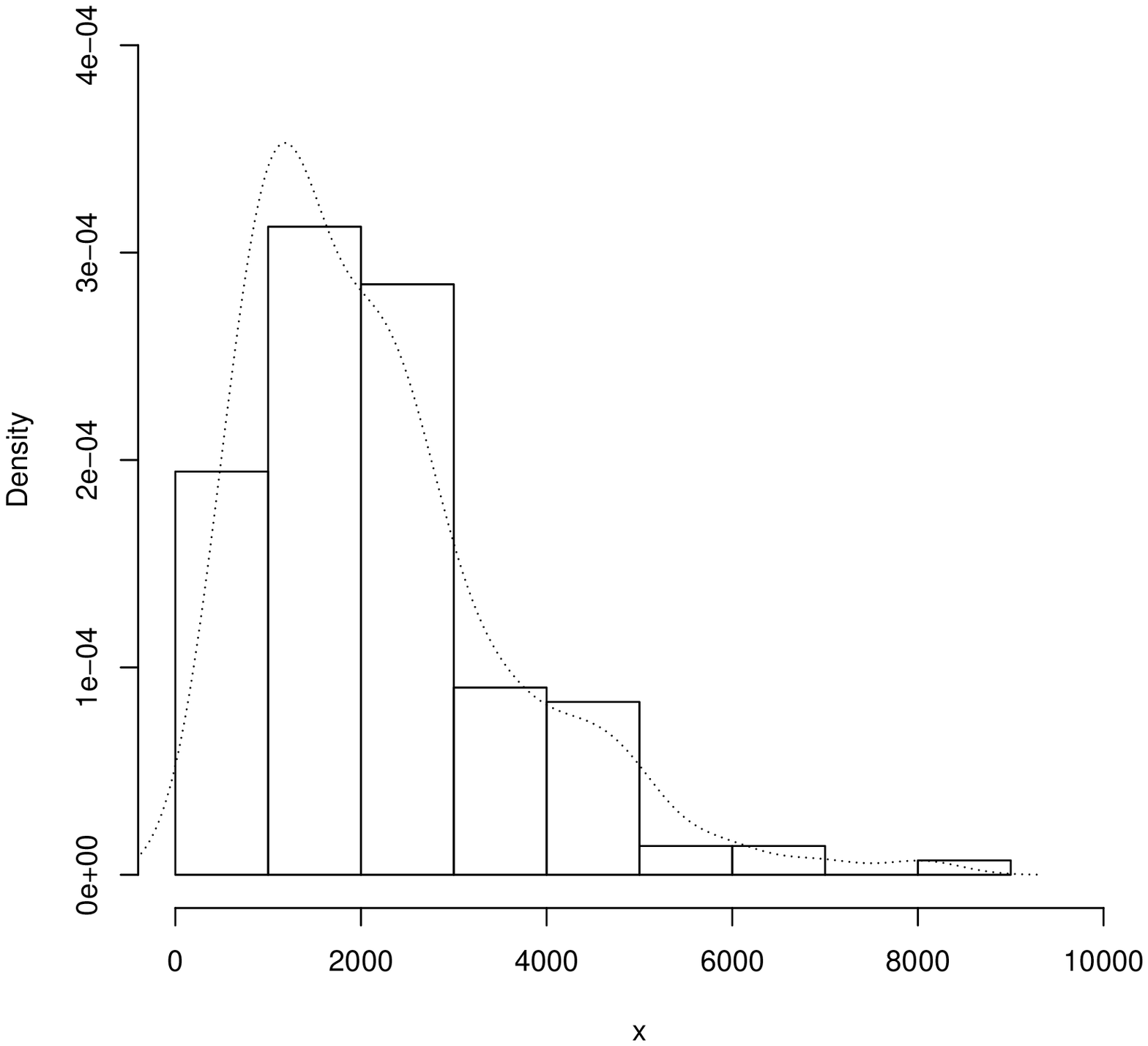}
	\end{center}
	\vspace{-1 cm}
  \caption{\rm Betaplasma data: Histogram of the amount of Beta-Carotene ($x$-variable).}\label{fig:histXbeta}
\end{figure}

\begin{table}
\begin{center}
\begin{tabular}{cccc}
\hline\hline
  & \textit{Gaussian} CWM & FMR  \\ \hline
$BIC$ & 4298.6  & 4324.5 \\
$\Lambda$ & 0.0934 & 0.2655   \\
$\cE$ & 70.38 & 54.11    \\
\hline\hline
\end{tabular}
\end{center}
\caption{\rm Betaplasma data:  values of BIC, $\Lambda$ and $\cE$ for both Gaussian CWM and FMR.}\label{tab:beta}
\end{table}

\begin{table}
\begin{center}
\begin{tabular}{cc|ccc} 
\hline\hline
{\em group} & {\em parameters} & {\em Gaussian CWM} & {\em FMR}  \\ \hline
1 & $\pi_1$ & 0.4036 & 0.3917  \\ 
  & $b_{10}$ & 103.0155 (\textit{2.2870}) & 89.9892 (\textit{1.0669})  \\
  & $b_{11}$ & 0.0087 (\textit{0.0011}) & 0.0032 (\textit{0.0004}) \\ 
  & & & & \\
2 & $\pi_2$ & 0.2789 & 0.3852  \\
  & $b_{20}$ & 436.1387 (\textit{5.1796}) & 125.4027 (\textit{1.4870}) \\
  & $b_{21}$ & -0.0393 (\textit{0.0014}) & 0.0301 (\textit{0.0006})  \\ 
& & & &  \\
3 & $\pi_3$ & 0.0283 & 0.1016\\
  & $b_{30}$ & 486.4574 \textit{(25.6135)} & -11.9203 (\textit{9.5508})  \\
  & $b_{31}$ & 0.1590 (\textit{0.0058}) & 0.2572 (\textit{0.0033})  \\ 
  & & & & \\
4 & $\pi_4$ & 0.2893 & 0.1215  \\ 
 & $b_{40}$ & 112.1419 (\textit{5.1502}) & 249.8431 (\textit{1.6601}) \\
& $b_{41}$ & 0.0524 (\textit{0.0051}) & 0.0419 (\textit{0.0006}) \\
\hline\hline \\
\end{tabular}
\end{center}
\caption{\rm Betaplasma: mixing weights and parameter estimates of the modes  for both Gaussian CWM and FMR. In parenthesis the standard errors 
of the estimtes are given.}\label{tab:parest_G4}
\end{table}

}\end{Ex}

\section{Concluding remarks}
\label{sec:concl}
In this paper, we presented a statistical analysis of Cluster-Weighted Modeling (CWM) based on elliptical distributions.  
Under the Gaussian case a detailed comparison among CWM and some competitive local statistical models such mixtures of distributions and mixtures of regression models. Moreover, based on both analytical and geometrical arguments, we have shown that CWM can be regarded as a generalization of such models. We remark also that Proposition \ref{prop:CWM-FMR} could be extended to {\em Finite Mixtures of Generalized Linear Models}. Further, our numerical simulations showed that CWM provides a very flexible and powerful framework in data classification, useful to perform a suitable data fitting. Moreover, with respect to the comparison with FMR and FMRC, \cite{Wedel:02} points out that assuming a distribution for the covariates makes inferences valid under repeated sampling and the model better suited to deal with general patterns of missing observations. 

In the second part of the paper,  we introduced new Cluster Weighted Modeling based on the Student-$t$ distribution for  robust clustering. In this context, in fitting noisy data, we considered a procedure for removing noise and estimating the parameters of the model on the remaining data following an approach  proposed in \cite{PeMcLa:00}. In this framework, recent literature on robust parameter estimation provides ideas for further research.

Another important issue, which  deserves attention for further research, concerns computational aspects of the parameter estimation in  CWM.
Parameters in CWM have been here estimated according to the maximum likelihood approach by means of the EM algorithm. In this paper, we have not presented  a detailed analysis of the behaviour of the EM algorithm under different conditions. However, our numerical simulations confirmed the findings of \cite{FaSo:10} in fitting mixtures of linear regressions and we point out that the initialization of the algorithms is quite critical. 
In our simulations the initial guess has been made according to either a preliminary clustering of data using a $k$-means algorithm or a random grouping of data, but our numerical studies
pointed out that there is no overwhelming strategy. Finally, we remark  that, in order to reduce such critical aspects,  suitable constraints on the eigenvalues of the covariance
matrices could be implemented, see e.g.  \citet{Ingr:04}, \citet{InRo:07}, \citet{GrIn:10}. This provides other ideas for future work.

\section*{Appendix: Decision surfaces of CWM}\label{sec:geom}
\begin{figure}[h] 
	\begin{center} 
		\includegraphics[height=7 cm, width=6 cm]{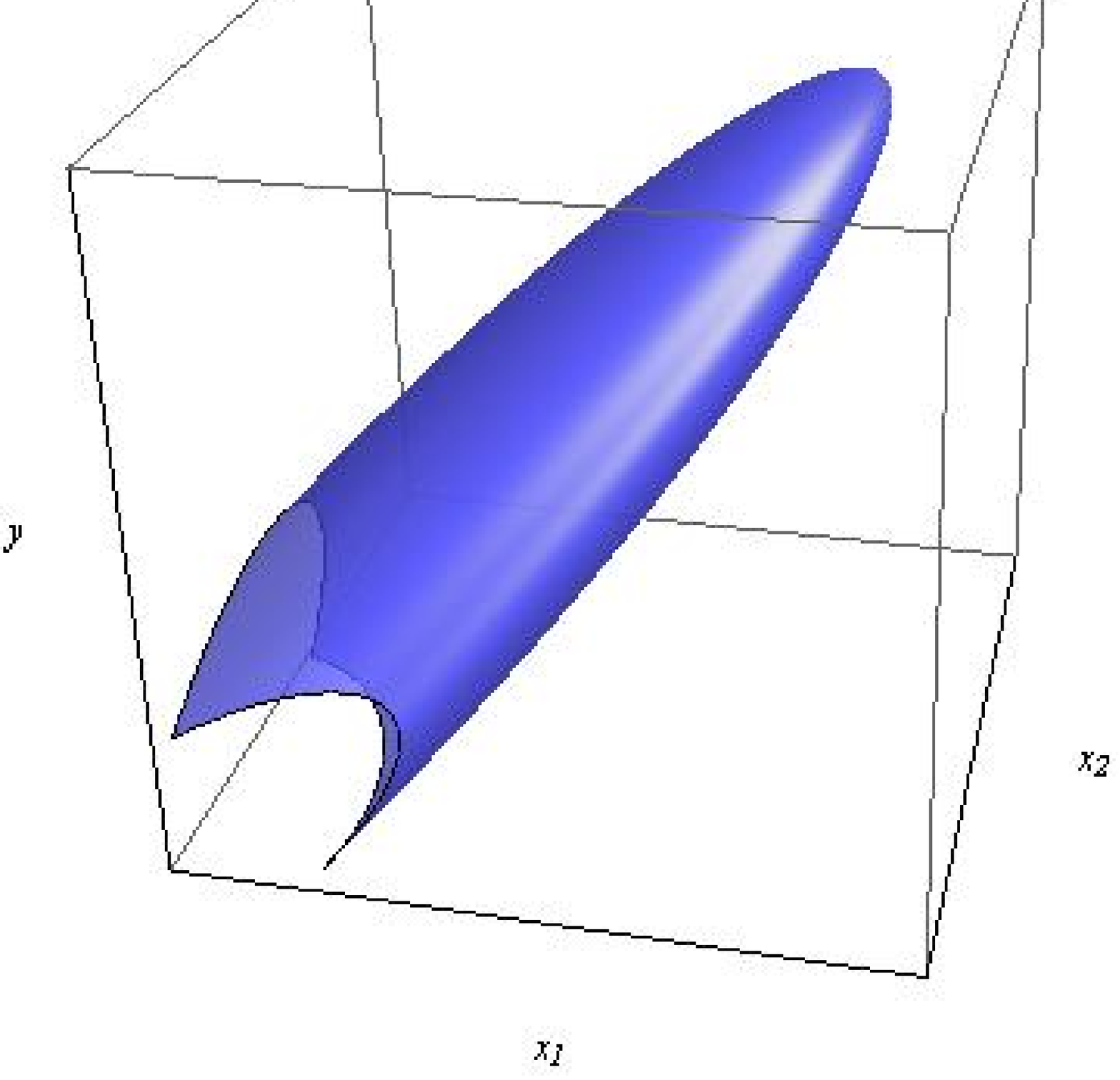}   
		\includegraphics[height=7cm, width=6 cm]{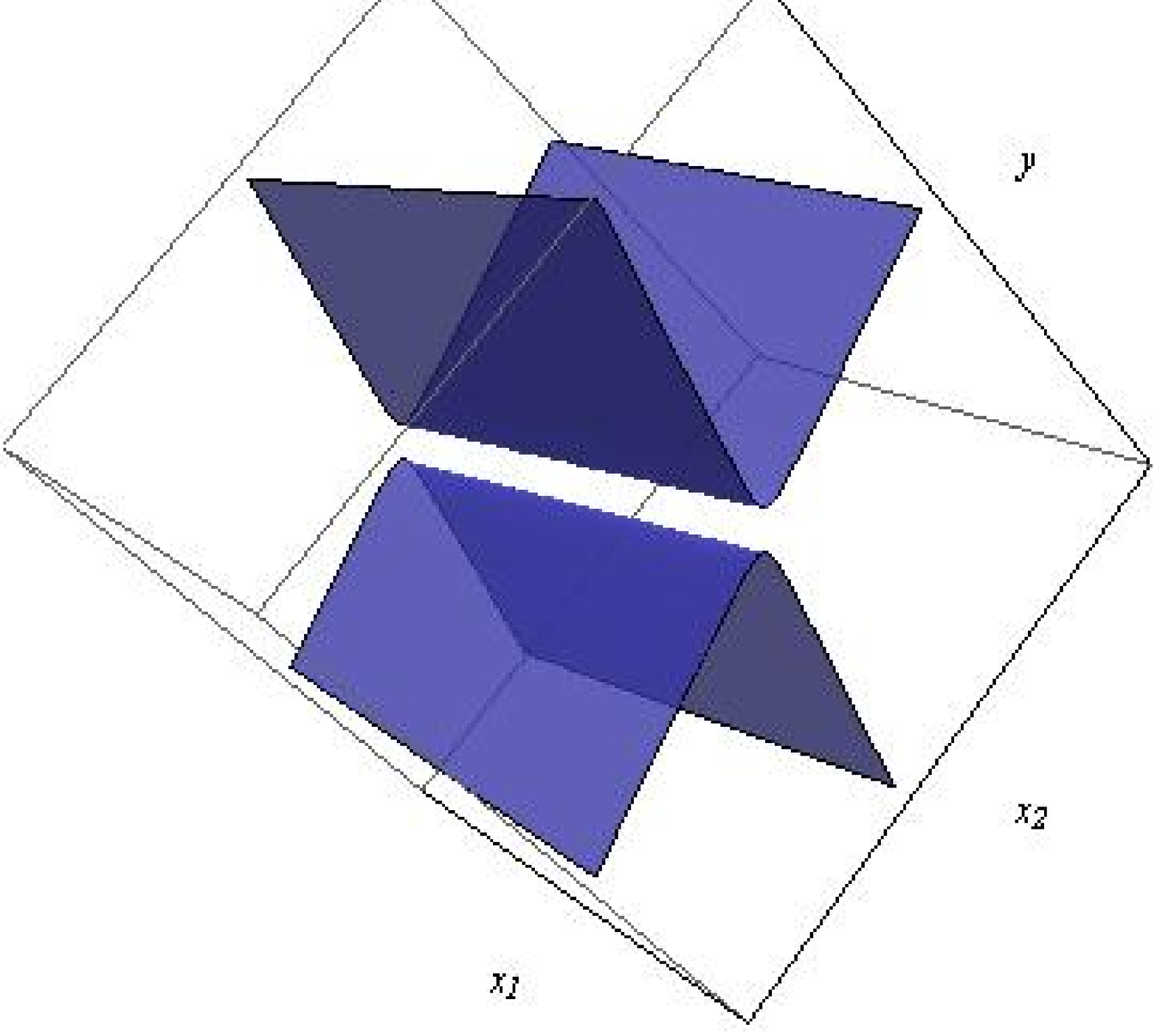}  
	\end{center}
 \vspace{-2 cm} 
 \caption{\textit {\rm Examples of decision surfaces for \textit{linear Gaussian} CWM (heteroscedastic case).}}\label{fig:GGCWM}
\end{figure}

\begin{figure}[h] 
	\begin{center}
		\includegraphics[height=8 cm, width=8 cm]{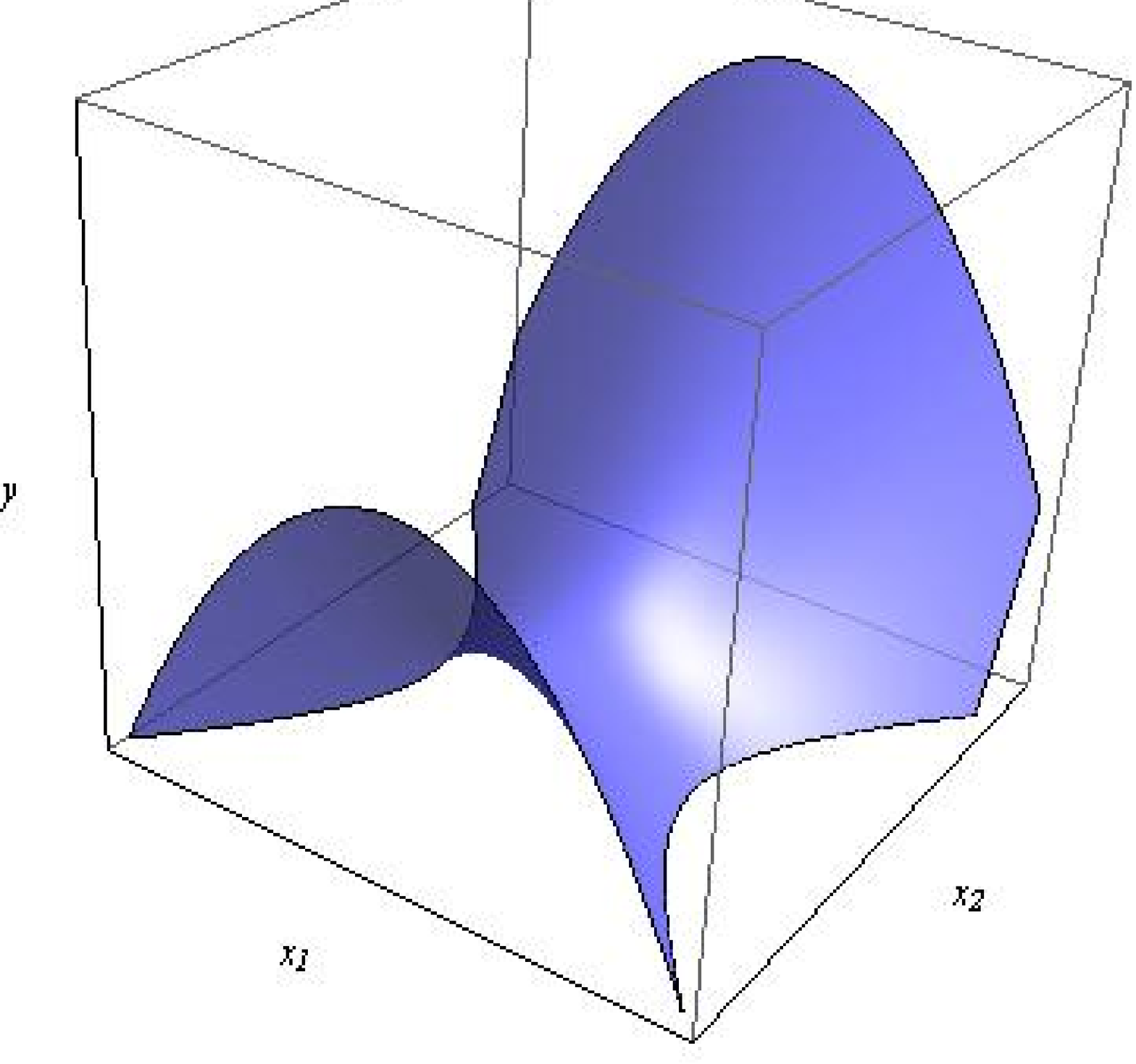} 
	\end{center}\vspace{-2 cm}
  \caption{\textit {\rm  Examples of decision surfaces for \textit{linear Gaussian} CWM (homoscedastic case). }}\label{fig:GGCWMh}
\end{figure}

The potentiality of CWM as a general framework can be illustrated also from a geometric point of view, by considering the decision surfaces which separate the clusters. In the following we will discuss the binary case,  
and in this case the decision surface is the set of $(\bx, y) \in \mR^{d+1}$ such that $p(\Omega_0|\bx,y)=p(\Omega_1|\bx,y)=0.5$. 
Given that $p(\bx|\Omega_g)\pi_g = p(\Omega_g|\bx) p(\bx)$, we can rewrite $p(\Omega_1|\bx,y)$ as:
\begin{align}
p(\Omega_1|\bx,y) & =\frac{p(y|\bx,\Omega_1)p(\Omega_1|\bx)}{p(y|\bx,\Omega_0)p(\Omega_0|\bx)+p(y|\bx,\Omega_1)p(\Omega_1|\bx)} = \frac{1}{1+\displaystyle{\frac{p(y|\bx,\Omega_0)p(\Omega_0|\bx)}{p(y|\bx,\Omega_1)p(\Omega_1|\bx)}}} \notag \\
& = \frac{1}{1+ \exp \left\{-\ln \displaystyle {\frac{p(y|\bx,\Omega_1)}{p(y|\bx,\Omega_0)} } -\ln \displaystyle {\frac{p(\Omega_1|\bx)}{p(\Omega_0|\bx)} }\right\} } \label{pO1|yx} \: .
\end{align}
Thus it results $p(\Omega_1|\bx,y)=0.5$ when
\begin{equation*} 
\ln \frac{p(y|\bx,\Omega_1)}{p(y|\bx,\Omega_0)} +\ln \frac{p(\Omega_1|\bx)}{p(\Omega_0|\bx)}   = 0 \: , \label{decsurf1}
\end{equation*}
which may be rewritten as:
\begin{equation} 
\ln \frac{p(y|\bx,\Omega_1)}{p(y|\bx,\Omega_0)} +\ln \frac{p(\bx|\Omega_1)}{p(\bx|\Omega_0)} + \ln \frac{\pi_1}{\pi_0} = 0 \: . \label{decsurf2}
\end{equation}
In the \textit{linear Gaussian} CWM, the first and the second term in (\ref{decsurf2}) are, respectively:
\begin{align*}
\ln \displaystyle {\frac{p(y|\bx,\Omega_1)}{p(y|\bx,\Omega_0)} }  & = \ln \frac{\sqrt{2 \pi \sigma_{\epsilon,0}^2}}{\sqrt{2 \pi \sigma_{\epsilon,1}^2}} + \frac{(y-\bb_0'\bx-b_{00})^2}{2 \sigma_{\epsilon,0}^2} -
\frac{(y-\bb_1'\bx-b_{10})^2}{2 \sigma_{\epsilon,1}^2} \:  \\
\ln \frac{p(\bx|\Omega_1) }{p(\bx|\Omega_0)} & = \frac{1}{2} \ln \frac{|\bSigma_0|}{|\bSigma_1|}+ \frac{1}{2} \left[(\bx -\bmu_0)' \bSigma_0^{-1} (\bx-\bmu_0)
- (\bx -\bmu_1)' \bSigma_1^{-1} (\bx-\bmu_1) \right] \: . 
\end{align*}
Then, equation (\ref{decsurf2}) is satisfied for $(\bx,y) \in \mR^{d+1}$ such that:
\begin{multline}
\ln \frac{\sigma_{\epsilon,0}}{ \sigma_{\epsilon,1}} + \frac{(y-\bb_0'\bx-b_{00})^2}{2 \sigma_{\epsilon,0}^2} -
\frac{(y-\bb_1'\bx-b_{10})^2}{2 \sigma_{\epsilon,1}^2}   +\frac{1}{2} \ln \frac{|\bSigma_0|}{|\bSigma_1|}+ \\ \frac{1}{2} \left[(\bx -\bmu_0)' \bSigma_0^{-1} (\bx-\bmu_0)
- (\bx -\bmu_1)' \bSigma_1^{-1} (\bx-\bmu_1) \right] + \ln \frac{\pi_1}{\pi_0}    =0 \: , \label{lnC0C1}
\end{multline}
which defines quadratic surfaces, i.e. {\em quadrics}. Examples of quadrics are spheres, circular cylinders, and circular cones. In Figure \ref{fig:GGCWM}, we give two examples of surfaces
generated by (\ref{lnC0C1}).
In the homoscedastic case $\bSigma_0=\bSigma_1=\bSigma$ it is well known that:
\begin{align}
\ln \frac{p(\bx|\Omega_1) }{p(\bx|\Omega_0)} & = \frac{1}{2} \left[(\bx -\bmu_0)' \bSigma^{-1} (\bx-\bmu_0) - (\bx -\bmu_1)' \bSigma^{-1} (\bx-\bmu_1) \right] \notag \\
& = \bw' \bx + w_0, \label{homsced}
\end{align}
where
\[ \bw = \bSigma^{-1}(\bmu_1-\bmu_0) \quad {\rm and} \quad w_0 = \frac{1}{2}(\bmu_0+\bmu_1)'\bSigma^{-1}(\bmu_0-\bmu_1)  .
\]
In this case, according to (\ref{homsced}), equation (\ref{decsurf2}) yields:
\begin{equation*} 
\ln \frac{p(y|\bx,\Omega_1)}{p(y|\bx,\Omega_0)} + \bw'\bx + w_0 + \ln \frac{\pi_1}{\pi_0} = 0 \: , \label{decsurf3}
\end{equation*}
see Figure \ref{fig:GGCWMh}.
%
%

As for  the \textit{linear} t-CWM, the first and the second term in (\ref{decsurf2}) are, respectively:
\begin{align*}
\ln \frac{p(\bx|\Omega_1) }{p(\bx|\Omega_0)}  & = 
 \ln \left[ \frac{\Gamma((\nu_1+q)/2) \Gamma(\nu_0/2)}{\Gamma((\nu_0+q)/2) \Gamma(\nu_1/2)} \right]  +\frac{1}{2} \ln \frac{|\bSigma_0|}{|\bSigma_1|}+ \\ &\quad  + \frac{\nu_0+q}{2} \ln 
\{\nu_0+ \delta(\bx, \bmu_0; \bSigma_0)\}- \frac{\nu_1+q}{2} \ln 
\{\nu_1+ \delta(\bx, \bmu_1; \bSigma_1)\} \: \\
\ln \displaystyle {\frac{p(y|\bx,\Omega_1)}{p(y|\bx,\Omega_0)} } & = \ln \left[ \frac{\Gamma((\zeta_1+1)/2) \Gamma(\zeta_0/2)}{\Gamma((\zeta_0+1)/2) \Gamma(\zeta_1/2)} \right] + \ln \frac{\sigma_{\epsilon,0}}{ \sigma_{\epsilon,1}} \\ &  \quad 
 + \frac{\zeta_0+1}{2} \ln \left[\zeta_0+ \left(\frac{(y-\bb_0'\bx-b_{00}}{ \sigma_{\epsilon,0}} \right)^2\right]-
\frac{\zeta_1+1}{2} \ln \left[\zeta_1+ \left(\frac{(y-\bb_1'\bx-b_{10})}{ \sigma_{\epsilon,1}} \right)^2\right] \:. \label{ln-py|xOt}
\end{align*}
Then, equation (\ref{decsurf2}) is satisfied for $(\bx,y) \in \mR^{d+1}$ such that:
\begin{multline}
 c(\nu_0, \nu_1, \zeta_0, \zeta_1)+ \ln \frac{\sigma_{\epsilon,0}}{ \sigma_{\epsilon,1}} 
 + \frac{\zeta_0+1}{2} \ln \left[\zeta_0+ \left( \frac{y-\bb_0'\bx-b_{00}}{ \sigma_{\epsilon,0}} \right)^2\right] + \\ -
\frac{\zeta_1+1}{2} \ln \left[\zeta_1+ \left( \frac{y-\bb_1'\bx-b_{10}}{ \sigma_{\epsilon,1}} \right)^2\right] + 
\frac{1}{2} \ln \frac{|\bSigma_0|}{|\bSigma_1|} \\  +  \frac{\nu_0+q}{2} \ln 
\{\nu_0+ \delta(\bx, \bmu_0; \bSigma_0)\}- \frac{\nu_1+q}{2} \ln 
\{\nu_1+ \delta(\bx, \bmu_1; \bSigma_1)\} + \ln \frac{\pi_1}{\pi_0} =0 \: , 
\end{multline}
where
\[ c(\nu_0, \nu_1, \zeta_0, \zeta_1) = \ln \left[ \frac{\Gamma((\zeta_1+1)/2) \Gamma(\zeta_0/2)}{\Gamma((\zeta_0+1)/2) \Gamma(\zeta_1/2)} \right] +
 \ln \left[ \frac{\Gamma((\nu_1+q)/2) \Gamma(\nu_0/2)}{\Gamma((\nu_0+q)/2) \Gamma(\nu_1/2)} \right] \: .
\]
We remark that, in this case, the decision surfaces are elliptical. 
\section*{Acknowledgements}
The authors sincerely thank the referees for their interesting comments and valuable suggestions. Thanks are also due to Antonio Punzo for helpful discussions.

\end{document}